# Synergies in offshore energy: a roadmap for the Danish sector

Feasibility study funded by the DHRTC

February 2021

Matteo D'Andrea[1], Mario Garzón González[2], Russell McKenna[1]
[1] School of Engineering, University of Aberdeen, UK
[2] DTU Management, Technical University of Denmark

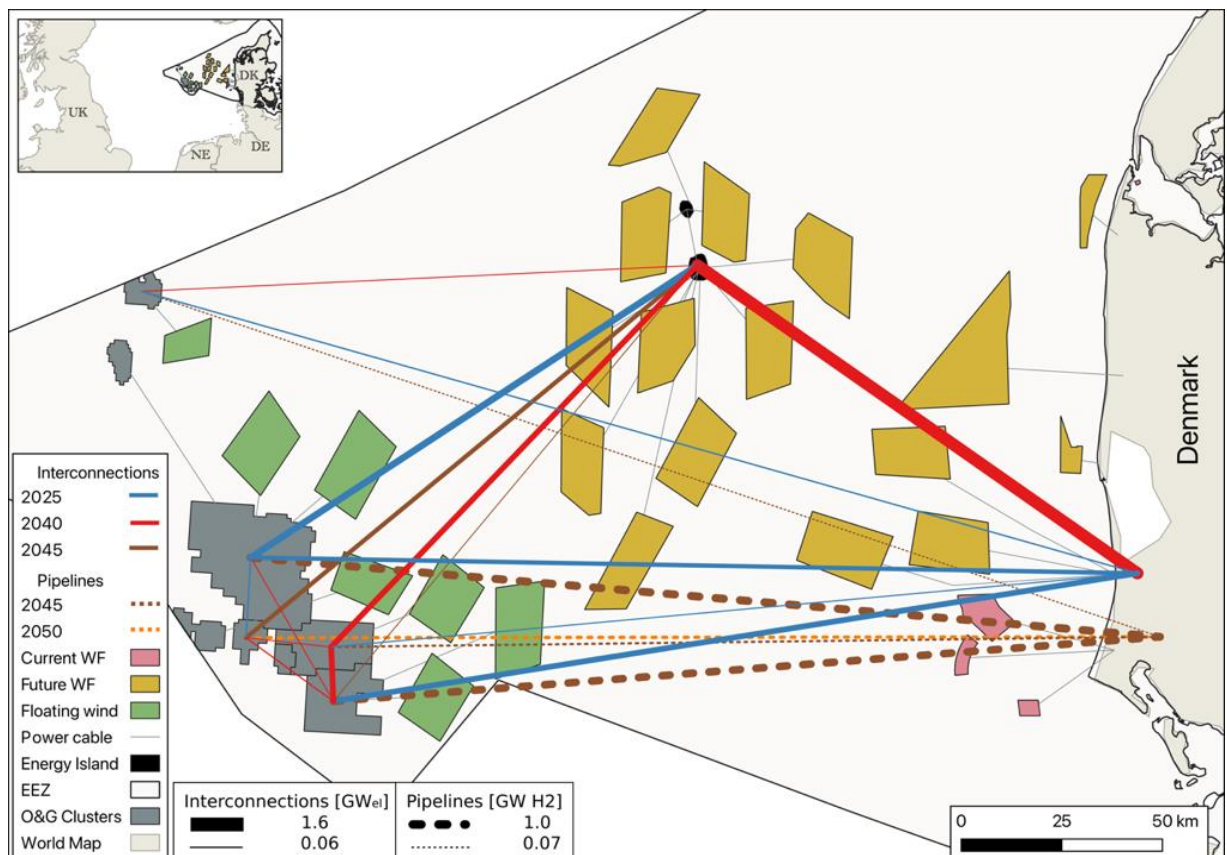

*Overview of Danish offshore region with developed energy system infrastructure to 2050*

# Executive Summary

This study sets out to analyze future synergies between the O&G and renewables sectors and explore how exploiting these synergies could lead to economic and environmental benefits. By firstly reviewing and highlighting relevant technologies and related projects, this report synthesizes the state of the art in offshore energy system integration, by focusing on three key technologies in electrification/floating wind, electrolysis/hydrogen production and Carbon Capture and Storage (CCS). Both existing O&G assets and these renewable technologies are techno-economically analysed with relevant technical and economic criteria. This results in an overview of the interconnected system of O&G assets within the DUC along with operational data and retirement schedules, as well as techno-economic characteristics of floating offshore wind, electrolysis and CCS technologies.

All of these preliminary results serve as input data for a holistic energy system analysis in the Balmorel modelling framework. With a timeframe out to 2050 and model scope including all North Sea neighbouring countries, this analysis explores a total of nine future scenarios for the North Sea energy system. The main results include an immediate electrification of all operational DUC platforms by linking them to the shore and/or a planned Danish energy island. These measures result in cost and $CO_2$ emissions savings compared to a BAU scenario of 72% and 85% respectively. When these platforms cease production, this is followed by the repurposing of the platforms into hydrogen generators with up to 3.6 GW of electrolysers and the development of up to 5.8 GW of floating wind. The generated hydrogen is assumed to power the future transport sector, and is delivered to shore in existing and/or new purpose-built pipelines. The contribution of the O&G sector to this hydrogen production amounts to around 19 TWh, which represents about 2% of total European hydrogen demand for transport in 2050. The levelized costs (LCOE) of producing this hydrogen in 2050 are around 4 $€_{2012}$/kg $H_2$, which is around twice those expected in similar studies. But this does not account for energy policies that may incentivize green hydrogen production in the future, which would serve to reduce this LCOE to a level that is more competitive with other sources.

The very short timeframe and small project team mean that the analysis presented here is relatively high-level, requiring many simplifying assumptions and resulting in some important technical details being overlooked and left for future research. In particular there remain significant uncertainties relating to the technical feasibility and future economic developments of most of the technologies analysed. For this reason, this report should be understood as a starting point for further and more detailed analysis, rather than a definitive roadmap for the sector. Given the explorative nature of this research, recommendations mainly relate to areas where future work should focus:

- The business case for the results reported here should be analysed from the operator's perspective, in order to provide a clear indication of possible value opportunities in repurposing existing assets.
- Collaborative research with offshore O&G engineers should clarify the technical constraints on electrification and repurposing.
- A more holistic energy system analysis should be performed in order to overcome some limitations in spatial resolution and technical approximations.
- A wider policy, regulation and market analysis needs to build on the above system analysis in order to assess the required framework conditions for a future integrated North Sea energy system.

# Table of Contents

# Table of Figures

# Table of Tables

# 1. Introduction

The Oil & Gas (O&G) sector is facing many challenges. These include international climate frameworks such as the Paris Agreement, consistently low crude oil prices together and the foreseeable End Of Life (EOL) of many O&G assets. On the one hand, there is need to clarify the implications of the wider energy transition on existing O&G operations. On the other hand, much offshore infrastructure is reaching or will soon reach (e.g. within 30 years) the end of its economically operational lifetime. The decision of the Danish Government in December 2020 to cancel new O&G explorations in the Danish North Sea means all existing production must have ceased by 2050. Faced with potentially very high decommissioning costs for this infrastructure, the O&G sector may benefit from a partial or complete repurposing of this infrastructure, during and/or after its operational lifetime.

This report presents the findings of a feasibility study about exploiting synergies between offshore O&G and renewable energy activities in the Danish Underground Consortium (DUC) region of the North Sea. The objective was to develop a roadmap for this sector out to 2050. The project was funded by the DHRTC alongside six other projects exploring the technical and economic feasibility of alternative uses for existing O&G infrastructure. The research was carried out from September 2020 to January 2021 between the School of Engineering, University of Aberdeen, and the Division of Sustainability, DTU Management, Technical University of Denmark. The very short timeframe and small project team mean that the analysis is relatively high-level, requiring many simplifying assumptions and resulting in some important technical details being overlooked and left for future research. For this reason, this report should be understood as a starting point for further and more detailed analysis, rather than a definitive roadmap for the sector.

This study consists of two main parts. The first performs a techno-economic analysis and screening of relevant coupling technologies between the two sectors, including offshore wind, hydrogen production with electrolysis and carbon capture and storage (CCS). The main outcomes from the implementation of the coupling technologies are: the extension of the platform's lifetime due to reduction in operational costs; the maximization of economic recovery of mineral resources from fields, (Peters et al. 2020); the reuse of the oil and gas infrastructure and the resulting reduction in decommissioning costs; as well as potentially saving space elsewhere and improving acceptance. The second part of the study employs the Balmorel open source energy system model from DTU Management in order to explore future scenarios for the DUC out to 2050. The model includes all European countries around the North Sea as well as some neighbouring states. It minimizes the costs of the whole energy system, including the O&G sector, to demonstrate possible future scenarios. The results show an immediate electrification of existing platforms in 2025 with connections to the shore and a Danish energy island. This is followed in 2035 onwards with a strong development of floating wind parks in combination with platform repurposing to produce hydrogen for the transport sector, which is transported to shore in existing or new pipelines.

This report is structured as follows. Section 2 gives an overview of the state of the art in the Danish O&G sector, with particular emphasis on decommissioning and energy system integration with renewables. Section 3 then performs a techno-economic analysis of the relevant O&G and renewable energy technologies. Section 4 subsequently performs a holistic energy system analysis of the whole DUC region out to 2050, before section 5 presents and discusses the results. The report closes with a summary and conclusions in section 6.

## 2. State of the art and related projects

This section gives an overview of the state of the art in the O&G sector and reviews some related projects that combine renewable technologies with O&G infrastructure.

### 2.1. Denmark's oil & gas history and future developments

Denmark has produced oil and gas since the 1960s. At the end of the sixties, after the discovery of the Groningen gas field in the Netherlands, the interest towards hydrocarbons in the North Sea rose. In 1962, Maersk obtained a 50 year long exclusive right to utilize the resources in the North Sea (DEA 2021a) . The same year, Maersk, Sheel and Gulf founded the Danish Underground Consortium (DUC).

The first discovery of oil was made at the Kraka field in 1966, while production started at the Dan field in 1972. In the following years, with the discovery of new fields, natural gas production started. Oil production exceeded consumption for the first time in 1991, thus Demark become self-sufficient in oil. The following years have been very positive for the O&G sector, with new discoveries leading to strong growth that peaked around 2004 with 22.6 million m3 of oil and 11.5 billion Nm3 of gas (DEA 2018a).

Figure 1 shows the oil production and the long term sales forecasts. The production grew until 2004, where it peaked and begun a consistent fall until now. Denmark was a net exporter of oil from about 1993 until 2018.

Overall, the oil production shows a decreasing trend. A more significant fall in the latest years is connected with the rebuilding of facilities on the Tyra field. Denmark is no longer expected to be a net exporter of oil in the future, except for in the year 2024, due to the expected production start-up of new developments.

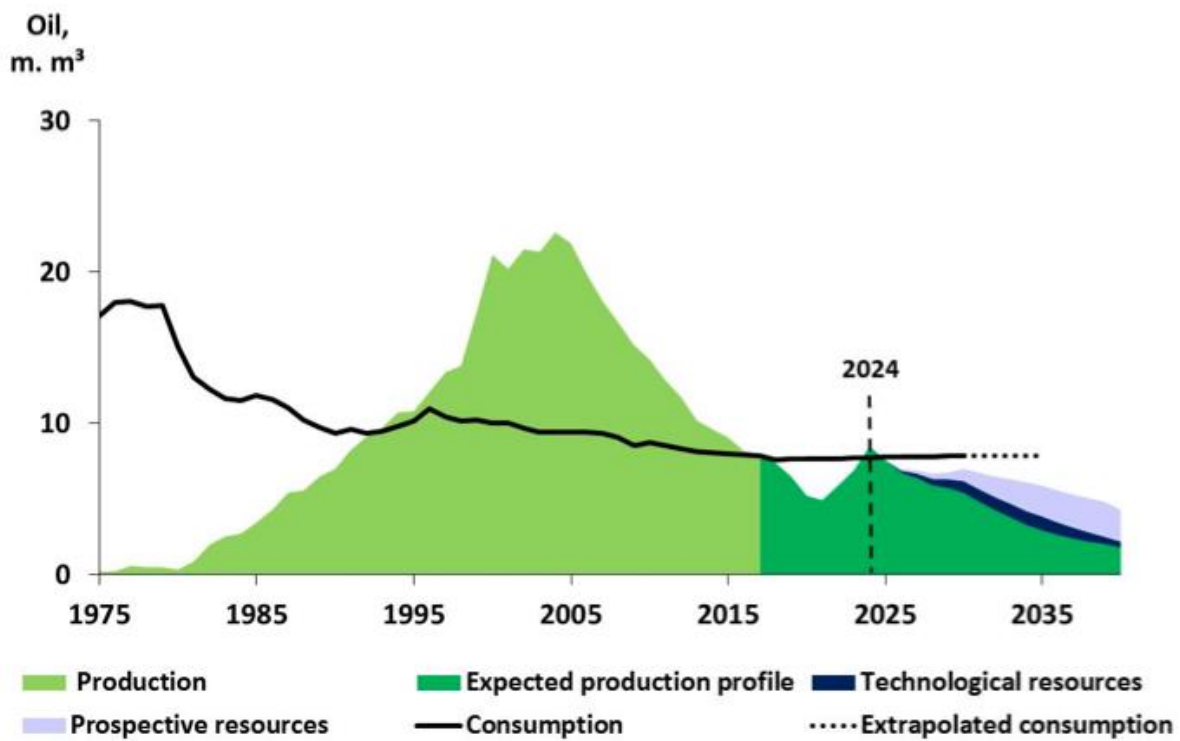

*Figure 1 Historical and long-term forecast for oil production and consumption* (DEA 2018a)

The gas production in Figure 2 shows a similar trend to the oil production. However, the production growth and fall are less steep and more consistent. Denmark is expected to be a net exporter of gas up to 2035, except for the years 2020 and 2021 due to the redevelopment of the facilities on the Tyra field.

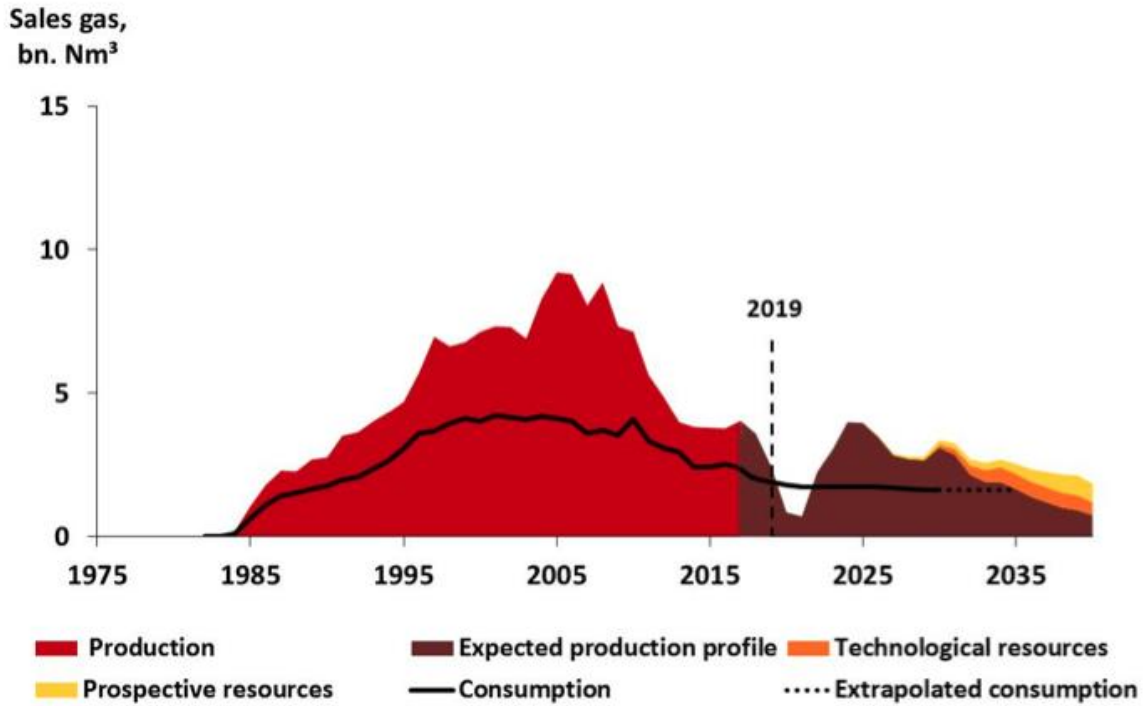

*Figure 2 Historical and long-term forecast for gas production and consumption* (DEA 2018a)

In December 2020, the Danish government voted in favour of cancelling all future licensing rounds for new O&G exploration and production permits in the Danish part of the North Sea. Further, all existing production must end by 2050 (Reuters 2020). The agreement still allows for a few new wells if they are connected to an already active drilling area try's plan to be carbon neutral by 2050, as defined in the Danish Climate Act (Ministry of Climate Energy and Utilities 2020).

This decision sets the end date of a decommissioning process that is expected to start much earlier. Large parts of the O&G infrastructure have been operational for many decades and the majority of the fields experience falling production. The advanced age of the installations means there is a high need for investments in maintenance and in certain cases for the establishment of new installations (Oil Gas Denmark 2017). Increasing costs for maintenance and operation might limit the time that production from certain facilities remains economically. There is a need for an overall assessment of the requirement for renovation and renewal, to create the best possible foundation for the continued optimization of production of remaining oil and gas in the subsurface (Oil Gas Denmark 2017). Several platforms are expected to be shut down and the production will be concentrated on the newest platforms such as the newly built facilities of the Tyra field. Towards 2050 and starting from the ones where production has ceased, all platforms will have to be decommissioned.

## 2.2. Decommissioning regulations and methods

In (DEA 2018b) the guidelines for the decommissioning of O&G facilities have been defined. These regulations set a tight schedule for the O&G operators, which have to decommission wells within 3 years from termination of operations, and subsea installations within 5 years of termination of operations. Further, the decommissioning plans must be submitted and approved two years in advance to the implementation of the decommissioning. Thus, it is necessary to define long-term planning horizons, with high uncertainty and high costs involved.

When it comes to decommissioning, there are multiple approaches that are summarized in Figure 3. At first, one can distinguish between the decommissioning of the jacket and the substructure. The former is often detached from the jackets and brought to shore in one piece, while the latter involves several decommissioning options. An overview of the decommissioning methods is presented in Figure 3.

The Danish Energy Agency requires foundations, framework poles or wells to be removed to and/or cut at a level below the seabed. Any decommissioning that consists of in-situ decommissioning, namely when all or parts of the installations are left on the seabed, must be justified by a Comparative Assessment of the selected and the alternative decommissioning methods (DEA 2018b).

An alternative to the full decommissioning of the O&G platforms is to reuse the existing offshore infrastructure. For example, the topside can be replaced with a new one that is placed on the existing jackets. Also, the existing gas pipelines could be refurbished to transport $CO_2$, hydrogen or alternative e-fuels. The reutilization of the O&G facilities for green projects, such as Power-to-X, can support the developments of new markets and business opportunities for the O&G sector, while contributing to the energy transition. Further details about the synergies between the existing gas infrastructure and renewable projects are provided in section 2.4.

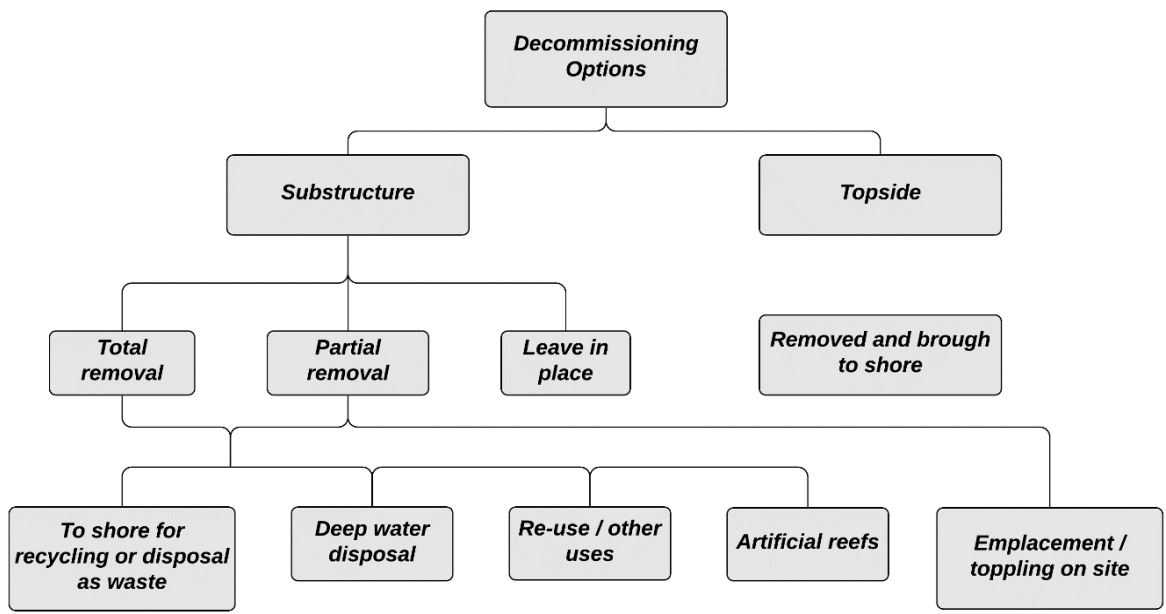

*Figure 3 Standard decommissioning options. Figure adapted from* (Leporini et al. 2019)

## 2.3. Decommissioning in the North Sea

The North Sea basin is home to 747 O&G offshore installations, excluding subsea stations (OSPAR 2017). The vast majority (603) are fixed steel, thus attached to the seabed by steel legs or built on concrete structures. The Continental Shelf of the UK (UKCS) counts the largest number of platforms (excluding subsea) in the North Sea with a share of 53% (417). Norway holds the second place with a share of 20% (144), followed by the Netherlands 15% (117) and Denmark 8% (61) (OSPAR 2017).

The North Sea counts many years of operation; production started in 1967 and peaked in the period from 1984 to 1993. The average age of the North Sea installations is 31 years, with the UKCS and Denmark being the oldest (30) and youngest (26), respectively.

To date, only 13% (94) of the North Sea installations had been decommissioned (OSPAR 2017). In (Decom North Sea 2014) from 2014, the platforms decommissioned are 88 of which only seven have had derogations granted to allow infrastructure to remain in place. Most of these are concrete gravity base infrastructure with single weights over 200,000 tonnes, except for one that is a fixed steel installation where the jacket and topside were removed and the footings of the jacket were left in situ. Furthermore, the study reports that, according to OSPAR and the North Sea regulators, 32 installations are likely to cease production and could be removed in the decade 2014-2024.

The sharing of knowledge between countries and companies in terms of platforms decommissioning can foster the market and strengthen the supply chain, resulting in a reduction of the costs. In particular, the Southern area of the UKCS, the Danish and the Dutch sectors have similar characteristics in terms of water depth, meteocean conditions and platforms.

UK and the Netherlands have defined precise strategies and methodologies for the future decommissioning of the O&G platforms. Both countries are learning from the experiences gained throughout the years to optimize the decommissioning process reducing costs and operational time. Denmark has removed the first platforms in 2019 for the redevelopment of the Tyra field. Therefore, Denmark can benefit from the experience of the neighbouring countries to define its decommissioning strategy.

### 2.3.1. The UK Continental Shelf (UKCS)

The UK has exploited its O&G reserves, primarily in the North Sea, since the 1960s. Production peaked in the mid-1980s and again in the late 1990s, and has declined steadily since then fields with 945 installations (713 operational, 137 closed down and 95 decommissioned) (OSPAR 2017).

The UK government has an objective to maximise the potential economic value of the UK's remaining O&G reserves, however, it expects that oil and gas will play a smaller role in meeting the demand for energy over time (NAO 2019). According to (Shell 2021), so far, around 10% of the UKCS has entered the decommissioning phase. Furthermore, currently some 40 decommissioning programmes have been submitted to the government's Department for Business, Energy and Industrial Strategy (BEIS) – the government body that regulates the decommissioning of offshore O&G installations and pipelines in the UK.

So far, operators have spent more than £1 billion on decommissioning in each year since 2014. The objective is to decommission one-fifth of the total well stock and 20% of the overall installations in the UKCS over the next decade (OGUK 2019). This consists of a removal of around 12 platforms per year up to 2025, with a peak of 24 in 2027.

The total cost of decommissioning UK offshore O&G production is estimated at £48 billion, which is expected to be reduced by 35% to reach the industry target of £39 billion (OGA 2020a). Over the next decade, the cumulative decommissioning expenditure is expected to be around £15 billion.

In line with the country's vision of net-carbon emission by 2050, UK's Oil & Gas Authority has started the Energy Integration Project in 2019, that explore how different offshore energy systems (oil and gas, renewables, hydrogen and carbon capture and storage) could be co-ordinated across the UKCS for environmental and efficiency gains. Within the ideas proposed in these projects, the use of decommissioned platform to reduce the decommissioning costs and create a new value for the platforms after the cease of production is evaluated (CoP) (OGUK 2019).

### 2.3.2. The Netherlands

Exploration and production of oil and gas in the Netherlands has been very successful over the last seven decades, specifically after the discovery of the giant Groningen gas field in 1959 (Herber and de Jager 2010). The natural gas resources as of 1st January 2020 are estimated at 172 billion Nm3, of which 18 billion Nm3 are in the Groningen gas field (Ministry of Economic Affairs and Climate Policy 2020). In 2019 the continental shelf produced 9.8 billion Nm3, 16.2% less compared to 2018.

The OSPAR database (OSPAR 2017) from 2017 reports 139 installation in the Dutch continental shelf has of which 118 operational, 21 closed down. The Netherlands is expecting to decommission a large portion of its O&G infrastructure over the next two decades. Furthermore, the decommissioning activities could intensify due to expected low O&G prices. For instance, according to (Rystad Energy 2019), the Netherlands recently announced that production at Groningen – Europe's largest gas field – will be halted in 2022, eight years earlier than initially planned. However, it is expected to have some residual production until 2030 as it is technically challenging to completely shut down production in such a short timeframe.

In 2016, Energie Beheer Nederland (EBN), an organization 100% owned by the Dutch state, published the "Netherlands Masterplan for Decommissioning and Re-use" which reports the current status of the O&G sector and defines the country's decommissioning strategy (EBN 2018). It confirms that the decommissioning process has started already; 23 out of ~150 offshore platforms have been removed and 200 km of 3500 km of pipeline on the Dutch Continental Shelf has been decommissioned The estimated costs for decommissioning are around 7 billion €2018, of which ~55% is related to offshore. The Dutch state bears approximately 70% of the costs.

The Dutch decommissioning strategy focuses on the reutilization of the O&G platforms for renewable investments. For instance, Nexstep, the Dutch association for decommissioning and reuse, focuses on stimulating the collaboration between key stakeholders to foster the re-use and decommissioning of O&G infrastructure. A pioneer project in terms of platform repurposing and hydrogen production offshore in the Dutch shelf is the PosHydon project presented in section 2.4.2.2.

## 2.4. Energy system integration of O&G and renewables

The North Sea has played an important role in the European energy system and will play a very important role in the future (North Sea Energy 2020). Two major trends are shaping the North Sea's energy landscape: on the one hand the decommissioning of the O&G facilities, and on the other hand the large investments in offshore renewable energy solutions. Unlocking the low-carbon energy potential of the North Sea requires integrated system thinking rather than merely sectoral optimization (North Sea Energy 2020).Therefore, it is important to identify synergies between the energy sectors and evaluate possible collaborations. Some opportunities for cooperation could be found between the O&G sector and the offshore wind sector, which are the largest players in the North Sea today (TNO 2018). The former's knowledge in terms of offshore operations, gas processing and subsurface technologies, combined with the large energy production at zero emissions and low cost obtained from the latter, could give the North Sea a key role in the energy transition. For instance, the power hub concepts are likely to be deployed in areas of mature O&G development (OGTC 2019). To assess the potential environmental impacts and support its realization the O&G sector can provide its expertise.

Sector coupling is a core instrument to exploit the synergies expressed above and achieve an optimal solution for the whole energy system both in terms of $CO_2$ emissions and cost savings (ENTSOE 2020). The coupling technologies evaluated in this report are the platform's electrification, Power-to-Hydrogen and CCS, as presented in the following sections.

### 2.4.1. Platform electrification
#### 2.4.1.1. Technology overview
The O&G platforms are located in remote areas of the Danish North Sea, at approximately 230 km from shore. The energy on the platform, required for gas processing, drilling, and daily activities, is provided by gas turbines. These are connected to electric generators, compressors and other auxiliary machinery. Smaller platforms are connected by cable or pipelines to receive energy or auxiliary fluids (e.g. water, compressed air). Security of supply is a crucial element; therefore, the gas turbines are required to run at all times and a redundancy criterion is applied to ensure power at all times. Since the energy consumption highly depends on the gas processing and the drilling activities which do not operate constantly, the gas turbines work at with variable load factors, hence at lower efficiencies than optimal. Furthermore, given the age of the gas turbines, the efficiency of the gas turbines is considered around 20%, with values as low as 13%.

The electrification of the platform would replace or decrease the use of the gas turbines and thereby reduce the overall energy consumption, $CO_2$ and $NO_x$ emissions. In addition, a cabled connection to an offshore power hub or to shore provides a reliable power supply, improving the security of supply. Moreover, an electrified platform could foster the research of other system integration concepts, such as carbon capture and storage and Power-to-Hydrogen (North Sea Energy 2020).

From a global point of view, the platform's electrification can help developing an offshore energy grid. Currently, the North Sea is home to several underwater cables that connect the countries around the North Sea. In (ENTSOE 2020) the need of improving the existing transmission corridors and the developing of new ones is stressed. This additional capacity allows for integration of renewable generation, by enabling cross-border exchanges and therefore minimizing curtailment. The offshore O&G platforms could act as energy hubs, hosting the grid facilities required to interconnect the countries.

The North Sea basin is home to multiple renewable energy projects that aim to harness clean energy such as wind turbines, wave and tidal energy converters and salinity gradient. In (SwAM 2019), ocean energy is considered an "untapped renewable energy source, despite decades of development efforts". Europe counts 21 tidal turbines (12 MW in total) and 13 wave energy converters (5 MW in total). These devices have not yet reached commercial viability. Ocean energy developers must still prove the reliability, availability and survivability of these technologies. The commercial phase is expected to be achieved around 2030 and the preferred site would be outside the Baltic Sea. The integration with the O&G sector is under research. A pilot project launched by the North Sea developer Chrysaor evaluates the use of Mocean's Blue Star wave energy converter (WEC) and EC-OG's HALO system to power subsea tiebacks and offshore oil field autonomous underwater vehicles (AUVs) (Recharge 2020). Regarding Denmark, Wave energy is estimated to be able to contribute 65% of the annual Danish energy consumption, according to Ocean Energy Systems (Ho et al. 2020). The Danish Energy Agency (DEA) refers to water power plants as "promising, but yet immature technology for renewable electricity". Denmark is hosting multiple demonstration plants at sea based on different concepts. The development work is focusing on development of technology concepts and conducting sea tests of large-scale models.

In the present study, the highest focus is on wind energy, as its technology is the most mature among the others. In relatively few years, offshore wind farms have become a reliable and competitive renewable energy technology. Currently, there are about 22 GW of offshore wind installed in Europe, of which 77% is located in the North Sea. According to (ENTSOE 2020), such capacity is expected to reach 70 GW by 2030 and 112 GW by 2040. In addition, the Green Deal indicates a potential need of more than 200 GW by 2050. Future wind farms will be placed farther offshore and into deeper waters due to both better, stable wind resources far from shore and the depletion of near-shore locations (Wind Europe 2019a). Therefore, future wind farms will be placed closer to the O&G platforms which are often far from shore as in the case of Denmark (see Figure 5) and so may increase the synergy opportunities. Furthermore, the participation of the O&G operators in the wind farm construction can support the wind farm sector development. (IEA 2019) stimates that around 40% of an offshore wind farm's lifetime costs, including construction and maintenance, have significant synergies with the offshore O&G sector. Furthermore, Power grid costs can be shared between O&G, wind farm, energy storage and transmission operators (OGTC 2020)

### 2.4.1.2. Hywind Tampen (Norway)
The Hywind Tampen project consists in partially electrifying five O&G platforms using a floating wind farm. It is a pioneering project since it is the world's first floating offshore wind farm supplying renewable power to offshore O&G installations and the largest offshore floating offshore wind farm to date (Equinor 2021).

The Snorre and Gullfaks offshore field facilities in Norway will be connected to an 88 MW floating wind farm. The farm consists of 11 wind turbines based on one of Equinor's floating offshore wind technologies, Hywind. The electricity produced is estimated to meet about 35% of the annual power demand of the platforms. From the reduction in the use of the gas turbines it is expected to offset 200,000 tonnes of $CO_2$ emissions per year, equivalent to emissions from 100,000 private cars. The project start up is expected in the third quarter of 2022.

### 2.4.2. Hydrogen production offshore
### 2.4.2.1. Technology overview
Electricity has a key role in the energy transition. Large investments in low-carbon electricity production coupled with the electrification of end-use sectors will result in a strong decrease of global $CO_2$ emissions. However, the deployment of large quantities of electricity requires large investments in grid expansion and a complex energy management system due to the high penetration of non-dispatchable energy sources. One possible solution is to decouple the energy production from its consumption. The Power-to-X concept embrace this idea by transforming electricity to other forms of energy that are easier to be stored such as methane, hydrogen and ammonia. It is important to stress that the electricity must have a low-carbon content to make the Power-to-X concept a sustainable solution.

A promising energy carrier is hydrogen; if produced from water electrolysis, which uses low-carbon electricity, it can decarbonise sectors that cannot be (fully) electrified, e.g. energy-intensive industries like iron and steel, chemicals, shipping and aviation. In addition, in the future, hydrogen is assumed to be stored at competitive costs in subsurface caverns of depleted gas fields (North Sea Energy 2020). [Therefore, within the Power-to-X concepts, Power-to-Hydrogen (P2H) has received a strong research attention from academia and industry sides.

A crucial role of the O&G sector within the P2H development comes from its positioning in the North Sea where, in the near future, the Danish North Sea area will host large capacities of offshore wind as discussed in section 3.2.1.

Furthermore, the O&G infrastructure such as pipelines and the platforms topside or jackets could be reused, reducing the capital expenditure of P2H. Moreover, depleted offshore O&G reservoirs may be used to store hydrogen. In terms of hydrogen transportation, the possibility of retrofitting the pipelines for transporting hydrogen in pure form or blended with natural gas is under research (Bonetto et al. 2019) . The latter method was proven to be feasible at volumetric percentage around 10% (Snam 2020). Finally, another driver for the implementation P2H offshore is the "not in my backyard" or NIMBY issue. The public acceptance regarding the offshore implementation of P2H could be higher resulting in a faster deployment of the technology.

### 2.4.2.2. PosHydon (the Netherlands)

The PosHydon project involves repurposing of an offshore platform that has ceased production, for the production of hydrogen via water electrolysis. The PosHydon pilot, the world's first offshore renewable hydrogen project, is an initiative of Nexstep, the Dutch association for decommissioning and reuse, and TNO, the Netherlands organisation for applied scientific research, in close collaboration with the industry (Neptune Energy 2021).

The Q13a is the first fully electrified platform in the Dutch North Sea, located approximately 13 kilometres off the coast of Scheveningen (The Hague). The electricity is provided from shore via cable and is generated from wind turbines offshore. Hydrogen is produced on the platform by water electrolysis. The electrolyser technology is the PEM (Proton Exchange Membrane) and its capacity is 1 MW (~200 $Nm^3$ $H_2$ /hr). A water treatment plant is required to demineralize the water fed to the electrolyser. This increases the area footprint of the hydrogen plant, which is a limitation for the implementation of large-scale projects. The aim of the pilot is to gain experience of integrating working energy systems at sea and the production of hydrogen in an offshore environment. This project helps save approximately 16,500 tonnes of $CO_2$ per year, equivalent to 80,000 private cars. The pilot plant is expected to be operational in 2021.

### 2.4.3. Carbon Capture and Storage (CCS) in depleted fields
### 2.4.3.1. Technology overview

Europe aims to reduce EU greenhouse gas emissions by at least 55% by 2030, compared to 1990 levels. A considerable share of the required $CO_2$ reduction is foreseen to be reached by implementing Carbon Capture and Storage (CCS). Storing $CO_2$ in depleted or depleting O&G fields has now been proven at a number of sites worldwide (Hannis et al.).

The deployment of CCS in the North Sea requires infrastructure for transport and storage. The O&G sector has the knowledge and the infrastructure for transport (e.g. compressors, pipelines) and storage of $CO_2$ (e.g. wells, reservoirs), thus it can foster the deployment of CCS in the North Sea. On top of that, the potential additional recovery of O&G by injecting $CO_2$ in hydrocarbon fields can increase the interest of O&G operators to implement CCS technologies (European Federation of Geologists 2020).

Currently, in Europe there are 13 commercial CCS facilities, and more than 11 commercial projects are targeting operation before 2030 (Global CCS institute 2020).

Offshore projects are being planned in several areas of the North Sea – the Goldeneye and Hewett gas fields in UK and the K12-B and P18-4 gas fields in the Netherlands (Hannis et al.).

BP, Eni, Equinor, National Grid, Shell and Total announced the formation of the Northern Endurance Partnership. With BP acting as Operator, the Group will develop offshore transport and storage infrastructure in the UK North Sea (Global CCS institute 2020). In Denmark, a $CO_2$ storage consortium formed by INEOS Oil & Gas Denmark and Wintershall Dea targets the development of $CO_2$ storage capacity offshore Denmark based on reusing discontinued offshore O&G fields for permanent CO2 storage.

A more detailed analysis about CCS is provided in section 3.2.4.

### 2.4.3.2. Project Greensand (Denmark)

The Greensand project aims at storing $CO_2$ in depleted gas reservoirs using existing O&G infrastructure. Project Greensand is an initiative of the Project Greensand consortium, led by INEOS Oil & Gas Denmark and partnered by Wintershall Dea and Maersk Drilling. The reservoir indicated for carbon storage is located at the Nini West field, in the Danish North Sea. DNV GL has issued a certification of feasibility for injecting 0.45 million tonnes $CO_2$ per year per well for a 10-year period (Maersk Drilling 2020). The $CO_2$ sequestration delivered by the project corresponds to 15-20% of the reduction required to reach Denmark's reduction target in 2030. Denmark's Energy Technology Development and Demonstration Program (EUDP) has awarded the project DKK 9.6 million (1.28 M€$_{2020}$) in support (Energy Watch 2020).

## 3. Techno-economic analysis

This section performs a techno-economic analysis of the two considered sectors to provide a basis for the energy system modelling in section 4. It begins with a characterisation of the O&G sector in section 3.1 before the relevant renewable energy technologies are analysed in section 3.2.

### 3.1. Oil and gas sector

In this section, the O&G offshore facilities are presented and described. Further, the techno-economic assumptions required to model these are provided.

### 3.1.1. Description of O&G platforms

The O&G offshore platforms have different infrastructure based on the operations carried out on the platform. Firstly, platforms can be divided between manned and unmanned. The latter are often called satellite platform since they are smaller and farther from the main facilities. Secondly, the platforms can be distinguished between the type of foundations and the operation performed (e.g. processing, flaring). An offshore platform typically consists of several key elements as displayed in Figure 4. These are:

- a topside, the above-water structure, where the offshore activities take place;
- a jacket, which is a steel structure that supports the topside. Alternatively, the foundations can be made of concrete or floating;
- footings, the heaviest section of the jacket, which anchors it to the seabed;
- pipelines for the export of oil and gas.

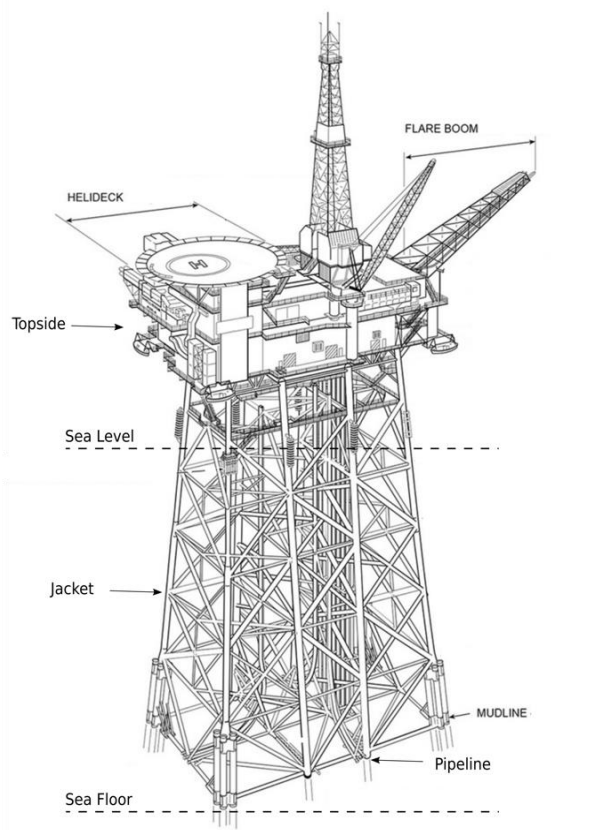

*Figure 4 Schematic of an offshore platform* (Lee Ramsden 2021)

### 3.1.2. Overview of the Danish O&G fields and infrastructure

Currently, the Danish O&G sector in the North Sea consists of 62 platform operating in 19 fields (OSPAR 2017). Total E&P DK is the operator in charge of production from 15 fields, while Ineos and Hess operate the remainder (DEA 2021a). In the present study, we focus on the 13 fields under the management of Total E&P DK and within the Danish Underground Consortium. These fields include 52 platforms, with data taken from the OSPAR database (OSPAR 2017). The name of the fields, together with the year of production start, the field age and the mean platform age within a field is provided in Table 1.

It can be noticed that some fields have been in operation for several decades. In some fields, the platforms have been built in different years, that is why the *mean platform age* (last column in Table 1) differs from the Field Age. The lifetime of a platform is expected to be between 20 and 30 years (Chakrabarti et al. 2005), thus one can notice that the Danish offshore facilities are close to the end of the expected lifetime; some have passed it already.

*Table 1 Fields under management of Total E&P DK and the DUC* (OSPAR 2017).

| Platform | First year of prod. | Field Age | Mean platform age |
|----------|---------------------|-----------|-------------------|
| Dan | 1972 | 48 | 36 |
| Gorm | 1981 | 39 | 37 |
| Skjold | 1982 | 38 | 30 |
| Tyra | 1984 | 36 | 31 |
| Rolf | 1986 | 34 | 34 |
| Dagmar | 1991 | 29 | 29 |
| Kraka | 1991 | 29 | 29 |
| Valdemar | 1993 | 27 | 18 |
| Roar | 1996 | 24 | 24 |
| Svend | 1996 | 24 | 24 |
| Harald | 1997 | 23 | 23 |
| Lulita | 1998 | 22 | 22 |
| Halfdan | 2000 | 20 | 15 |

The platforms are located in the western area of the Danish North Sea, at around 230 km from shore. A map of the area is presented in Figure 5, from which it is clear that the majority of the fields are close together. This area, called *Doggerbank*, is a large sandbank with shallow waters (~35 m deep) that is rich in hydrocarbons. Oil and gas are transported to shore via pipelines which are connected to Tyra, Gorm and Harald facilities. The pipelines' landing point is in the area of *Nybro* for gas and *Fredericia* for oil.

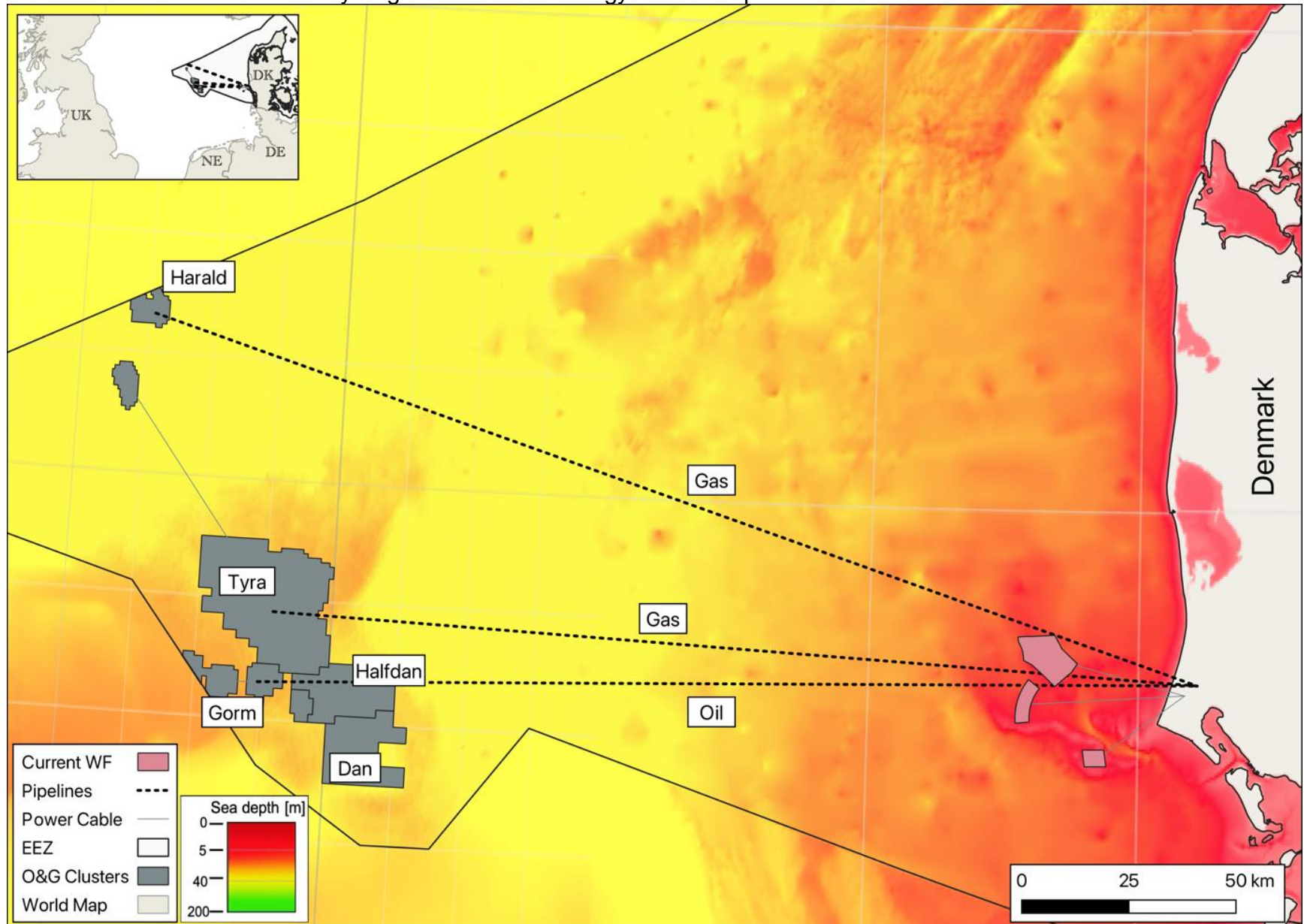

*Figure 5 O&G fields under operation of the DUC and TOTAL E&P DK. The colour heatmap shows the water depth.* (4C Offshore Database 2020; DEA 2021a).

### 3.1.3. Platform clustering

The platforms have been aggregated according to the field they are operating on. Further, small fields have been aggregated to larger ones which receive and process the O&G production. The cluster name refers to the main field within the aggregated ones. Five clusters of platforms have been identified as presented in Table 2.

*Table 2 Clusters of platforms with the fields assigned to each and the resulting number of platforms*

| Cluster Name | Number of platforms | Fields assigned to the cluster |
|---|---|---|
| Tyra | 18 | Tyra, Valdemar, Svend, Roar |
| Gorm | 11 | Gorm, Rolf, Dagmar, Skjold |
| Dan | 13 | Dan, Kraka |
| Halfdan | 8 | Halfdan |
| Harald | 2 | Harald, Lulita |

An example of a cluster of platforms is presented in Figure 6 for the Tyra field. The Tyra field is one of the largest; it is divided into areas (e.g. East, West, etc.) and has multiple processing, accommodation and satellite platforms. Satellite platforms are located in neighbouring fields of Tyra (i.e. Svend, Valdemar and Roar).

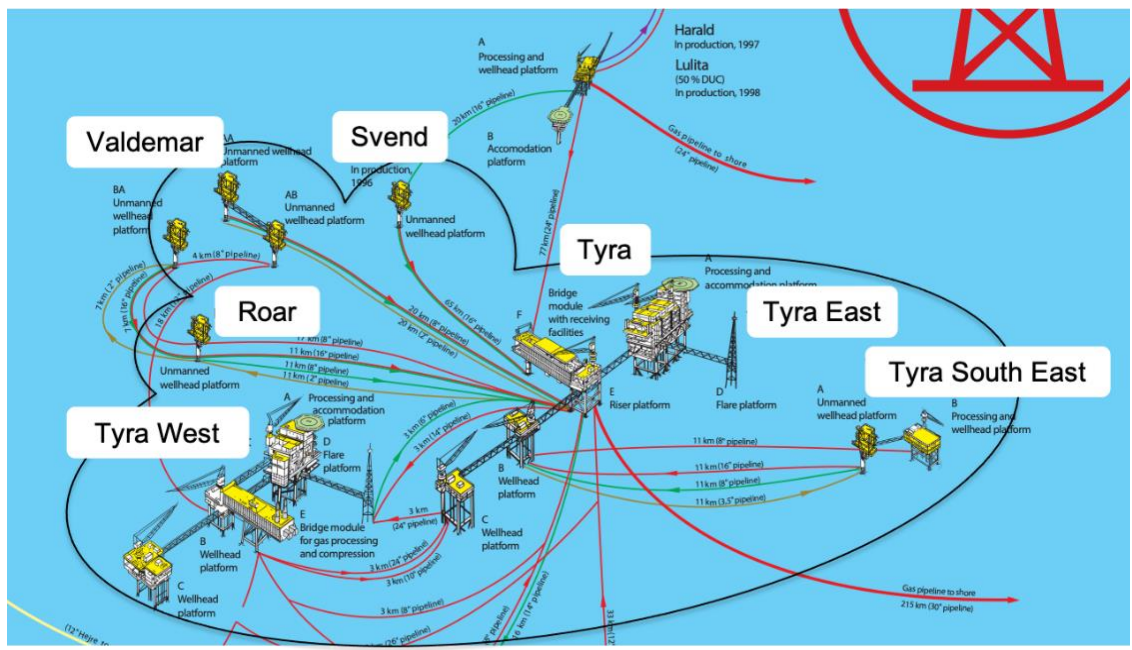

*Figure 6 Tyra's cluster of platforms*

The clustering methodology aggregates all platforms within a field to form a (hypothetical) single large platform that takes the name of the field. The characteristics of the individual platforms (e.g. energy consumption, weight, etc.) are aggregated together. The assumptions considered are the following:

- The pipelines within each cluster are neglected.
- Each cluster has an electricity demand which is the sum of the demand of each platform.
- The decommission of a cluster consists of ceasing operations of all platform within the cluster at once.
- The aggregated platform weight of a cluster is the sum of the platforms' weights within the cluster.

- The gas turbine's capacity of a cluster is the sum of the turbines of the individual platforms within the cluster.
- $CO_2$ emissions of a cluster are the sum of the emissions from the individual platforms within the cluster.

### 3.1.4. Decommissioning process and timeline

When the O&G production of a field ceases, the platforms operating there are decommissioned. In this study, the decommissioning process can be of two kinds. In the first, the platform topside and whole substructure are removed and brought to shore (conventional decommissioning). Alternatively, the platform's topside is removed and replaced with a new one or refurbished. The latter method assumes that the platform is going to be used for other purposes such as P2H and/or CCS. Moreover, the platform's jackets are assumed able to support a new topside, an assumption returned to in the discussion.

   The decommissioning timeline is presented in Table 3, based on the platforms' age. Specifically, it considers the mean cluster age and the age of specific platforms in the cluster. For example, in Table 3 one can notice that the mean cluster age of Dan is higher than Harald's, however, Dan ceases operations after Harald. The reason behind this choice lies in the age of the individual platforms in the clusters. In the Dan field, there are two platforms, namely Dan FG and Dan FF, with age 15 and 23, respectively, while Harald has two platforms of 23 years old. Moreover, the Dan field has larger reservoirs than Harald, thus it is considered to produce for a longer period.

   Regarding Tyra's facilities, since these are currently under renovation, it is assumed to be operational until 2042. In this year the concession for drilling given by the Government to the DUC expires and so all operations are assumed to cease. The cease of production year represents the last year with an energy demand on the platform and represents the beginning of decommissioning process, which is assumed to last 5 years. For example, a platform that is decommissioned in 2025 can be available for alternative uses from 2030 onwards.

*Table 3 Clusters' decommissioning timeline. *The mean age of Tyra's cluster doesn't consider the latest renovation on the field.* (OSPAR 2017)

| Cluster | Start of production year | Mean cluster age | Assumed cease of production year |
|---------|--------------------------|------------------|----------------------------------|
| Gorm | 1981 | 36 | 2025 |
| Harald | 1997 | 23 | 2030 |
| Dan | 1972 | 34 | 2035 |
| Halfdan | 2000 | 15 | 2040 |
| Tyra | 1984 | 28* | 2042 |

### 3.1.5. Energy system and $CO_2$ emissions on the platforms

The energy on the platform, required for gas processing, drilling, and daily activities, is provided by gas turbines. These are generally located on the manned platforms, while satellite platforms receive electricity and fluids via cables or pipelines.

   (DEA 2021a) provides a breakdown of the O&G production in different categories such as fuel, flaring, injection and export. The energy demand on the platform is derived from the annual natural gas consumption for fuel purposes on the field. The energy consumption on the platform is strongly influenced by the volume of gas processed, which fluctuates due to the maturity of the reservoirs. Therefore, to have a more consistent estimation of the platforms' energy demand, the mean yearly consumption of the last 5 years is considered. Then, based on the assumption of a constant demand, the mean hourly energy demand is obtained. The result is provided on a cluster level in Table 5. The turbines capacities for each platform are reported in the Environmental and Social Impact Statement (ESIS) of the field (DEA 2021a). Table 4 presents the turbine's capacities on a cluster level.

In addition, the heat demand on the platform has not been modelled. It is assumed that if the platform is electrified, the heat required for processing oil and gas can be produced from electricity. This is returned to in the discussion.

*Table 4 Clusters' installed capacity and natural gas consumption for fuel purposes* (DEA 2021a)

| Cluster | Total gas turbine capacity [MW] | Natural Gas consumption for fuel purposes [mill Nm$^3$] |
|---|---|---|
| Dan | 453 | 148 |
| Gorm | 311 | 76 |
| Tyra | 255 | 155 |
| Harald | 50 | 14 |
| Halfdan | 164 | 70 |
| | **1233** | **463** |

The greenhouse gas emissions on the platform are assumed to be related only to the use of natural gas. Additional emissions specific to the drilling activities are neglected. Natural gas is consumed in the gas turbines and in the gas flaring and venting process: "Gas flaring is the process of burning-off associated gas from wells, hydrocarbon processing plants or refineries, either as a means of disposal or as a safety measure to relieve pressure" (Emam 2015). Venting is a similar process which does not involve any combustion but consists of releasing the natural gas in the atmosphere. The volume of gas vented is generally very low – 0.05% of all produced gas in the UKCS (OGA 2020b) – thus it is neglected in the present study. The consumption of the natural gas for fuel and flaring purposes, provided by the Danish Energy Agency (DEA 2021a), is presented in Table 5. Natural gas has an emission factor of 2.28 tCO$_2$/1000 Nm$^3$ (Nielsen et al. 2010) which is used to calculate the CO$_2$ emissions related to the gas usage.

*Table 5 Natural Gas consumption and related CO2 emissions. The emission factor of natural gas is 2.28 tCO2/1000 Nm$^3$.* (Nielsen et al. 2010; DEA 2021a)

| | Natural gas consumption | | | CO$_2$ emissions | | |
|---|---|---|---|---|---|---|
| Cluster | Total [M Nm3] | For fuel purposes [M Nm$^3$] | For flaring purposes [M Nm$^3$] | Total [kt CO$_2$] | From fuel purposes [kt CO$_2$] | From flaring purposes [kt CO$_2$] |
| Dan | 205 | 148 | 17 | 468 | 429 | 39 |
| Gorm | 114 | 76 | 33 | 260 | 185 | 75 |
| Tyra | 177 | 155 | 19 | 404 | 361 | 43 |
| Harald | 19 | 14 | 2 | 43 | 39 | 5 |
| Halfdan | 80 | 70 | 6 | 183 | 169 | 14 |
| **Total** | **595** | **463** | **77** | **1359** | **1183** | **176** |

### 3.1.6. Integration of renewable energy technologies on the platform

The gross available area on the platform can determine whether a technology such as a P2H plant can be placed on the existing platform and what would be the maximum capacity that is feasible to install. When the platform is repurposed for alternative uses, two options are generally considered. The platform's topside can be renovated to make space for new machines, or it can be substituted with a new one. Generally, the topside is built onshore as a single block, therefore, the machines, such as the gas turbines, are strongly integrated into the structure and the removal of these machines might not be feasible. However, the platform can have some available space – usually on the top level – which can be used without any intervention. The topside replacement is therefore often the preferred solution, as was done for Tyra's redevelopment (Total Denmark 2021). It reduces the expensive and dangerous operations offshore and allows to design a platform that integrates renewable energy plants such as P2H and CCS.

Another criteria for the integration of a renewable plant is to not exceed the maximum weight allowance of the jackets. However, (Catrinus and Jepma 2017) demonstrates that the availability of space is a tighter bound than the weight allowance, when integrating renewable energy plants on an existing platform. This is returned to in the discussion.

In this study, it is assumed that for each cluster of platforms, 50% of the platforms may have the topside replaced by a new one with the renewable energy plant embedded. Moreover, it is assumed that the new topside has the same weight of the old one and the jackets will be able to support it. This assumption of a new topside allows more flexibility about the limitation of the available space. However, due to the uncertainty in the sizes of the renewable energy plants such as P2H and CCS, it was decided to neglect any limitation in the maximum capacity that can be installed on the platform. In this regards, it is important to consider that the platforms have been aggregated so the capacity of a hydrogen plant, for instance, would be shared among the platforms within the cluster.

### 3.1.7. O&G cost assumptions

**Decommissioning costs** are taken from (IMSA Amsterdam 2011), which employs the decommissioning costs of the UK to assess the decommissioning cost of other European O&G sectors. The study identifies a linear cost of decommissioning of 0.015 $M€_{2011}$/ton. In their estimation, the cost of well plugging, and abandonment is not included. The latest UK decommissioning cost estimation (OGA 2020a) considers this operation to account for 45% of the total decommissioning costs. This results in a cost of decommissioning of 0.025 $M€_{2012}$/ton. The total decommissioning cost for each cluster is found as the product of the linear decommissioning cost and the cluster's aggregated weight. The decommissioning cost of each cluster is provided in Table 6.

*Table 6 Clusters' weights, decommissioning costs and operational costs* (OSPAR 2017)

| Cluster | Total Weight [kt] | Topside Weight [kt] | Decommissioning costs [M€_{2012}] | OPEX [M€_{2012}/year] | OPEX repurposed [M€_{2012}/year] |
|---------|-------------------|---------------------|-----------------------------------|-----------------------|----------------------------------|
| Dan | 60 | 43 | 1678 | 243 | 24 |
| Gorm | 49 | 34 | 1378 | 200 | 20 |
| Tyra | 68 | 45 | 1911 | 277 | 28 |
| Harald | 13 | 7 | 358 | 52 | 5 |
| Halfdan | 37 | 20 | 1028 | 149 | 15 |
| | **227** | **150** | **6353** | **921** | **92** |

**Operational and maintenance costs (OPEX)** of a platform consists in the expenses required to keep the platform operational. The total annual OPEX of the Danish O&G sector are provided by the Danish Energy Agency (DEA 2021a). From an analysis of the OPEX of the platforms in the UKCS, a linear trend between the platform's weight and the platform's OPEX is observed. Therefore, the linear OPEX is calculated as 0.004 M€$_{2012}$/Ton per year. The cluster's OPEX is found as the product of the linear OPEX and the cluster's aggregated weight. The clusters' OPEX values are also provided in Table 6.

   **Costs of repurposing the platform** involve the decommissioning of the topside in 50% of the platforms and the construction of new ones. On the other hand, it is assumed that the wells, the jackets, the pipelines and the subsea structure can be used as they are, hence, the decommissioning costs of these structures are saved. The cost of decommissioning the topside is assumed to be about 30% of the full decommissioning cost of the platform. This percentage is based on the breakdown of the UKCS decommissioning costs provided Figure 7. It is considered that decommissioning the topside includes parts of the costs of removals, recycling, site monitoring and owner's costs. The decommissioning costs are calculated as the product of the weight of the cluster, the linear cost of decommissioning and the factors previously mentioned (i.e. 50% and 30%). The cost of building a new topside is about 40 €$_{2017}$ /kg (Catrinus and Jepma 2017). The cost of a new topside is therefore calculated as the aggregated weight of the topsides of the cluster times the unitary cost of building the topside aforementioned multiplied by 0.50, since we consider only half of the platforms in a cluster. Furthermore, the operational expenses of the repurposed cluster are assumed to be 10% of the cluster's OPEX when producing oil and gas.

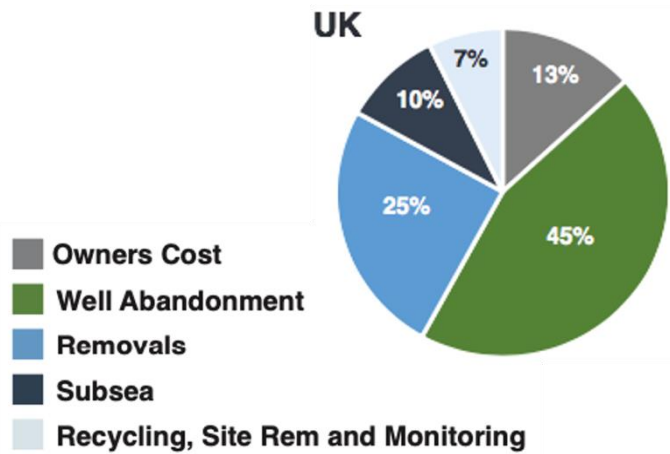

*Figure 7 Share of decommissioning costs estimated for the UK Continental Shelf* (OGA 2020a)

## 3.2. Renewable technologies integrated with the O&G sector

This section presents the techno-economic characteristics of the renewable technologies that can be integrated with the O&G sector, with a focus on offshore wind turbines (section 3.2.1), hydrogen (3.2.3) and carbon capture and storage (3.2.4).

### 3.2.1. Offshore wind energy status and plans in Denmark

Denmark has 14 offshore wind farms with a cumulative capacity of 1.7 GW, with 825 MW in the North Sea area that consists of the spatial boundaries of this analysis (Wind Europe 2019a). Currently, most wind farms are located near-shore at a distance between 15 and 40 km and sea depth between 5-20 m. On behalf of the Danish Energy

Agency and in cooperation with Energinet, COWI carried out a fine screening of the optimum sites for Denmark's new planned offshore turbines (4C Offshore Database 2020; COWI 2020a, 2020b).The screening identified areas where it is possible to establish the offshore wind farms with a capacity of more than 18 GW, of which 14 are located in the Danish North Sea. The development area identified are Nørdsoen I, which includes Thor and Nordsøen II+III+Vest that will be connected to the energy island. Furthremore, Vattenfall is building two wind farms near shore, Vesterhav Nørre and Vesterhav Syd with a cumulative capacity of 350 MW (4C Offshore Database 2020). The techno-economic data and the geographical location of these wind farms are provided in Table 7 and Figure 8, respectively.

Future wind development areas are located farther from shore due to the scarcity of near-shore locations and the need of reducing the visual impact of future wind turbines, that are increasing in size – the rated capacity of turbines installed in 2019 is 7.8 MW, 1 MW larger than last year (Wind Europe 2019a). A positive characteristic of the Danish area of the North Sea is that the seabed depth has a low correlation with the distance from shore. Thus, locations farther from shore have similar costs than closer ones, as shown in the last column of Table 7. The total planned capacity in the North Sea area is about 14.3 GW, which represents an increase of 1700% compared to the current installed capacity of 0.8 GW. If we assume the last wind farms is completed in 2050, it is required to install 450 MW each year.

Further, in the Climate Agreement of 22 June 2020 (Ministry of Climate Energy and Utilities 2020) the building of two energy islands was approved. The vision involves a network of hubs (energy islands) that connect far offshore wind farms to North Sea Countries' energy markets. At first, the energy generated from the surrounding wind farms is converted on the island into direct current electricity, before being exported by a series of interconnectors to the linked North Sea countries. In the long term, the intention is for the energy islands to be connected to additional wind farms and serve other purposes such as energy storage and conversion to other energy sources. (COWI 2020a, 2020b) describes two energy islands located at 100 km from the Danish coast. At first, a capacity of 3 GW offshore wind will be connected to the island, of which 2 GW are connected to shore and 1 GW is transported to Holland. The aim is to connect up to 10 GW to the island. The costs of building the two energy islands are included in the Capex of Nordsøen II+III+Vest wind farms Table 7. This island costs corresponds to around 1,500 M€$_{2012}$ when 10 GW are connected to it.

*Table 7 Wind farms techno-economic data. *Costs of Thor wind farm were estimated with the costs of Nordsøen I.* (4C Offshore Database 2020; COWI 2020a, 2020b)

| Wind Farm | Capacity [MW] | Production year | Connected to | Foundation Type | Capex [M€$_{2012}$/ year] | Opex [M€$_{2012}$/ year] | LCOE [€$_{2012}$/ MWh] |
|---|---|---|---|---|---|---|---|
| Vesterhav Nørre | 180 | 2021 | Land | Fixed-bottom | - | - | 63.6 |
| Vesterhav Syd | 170 | 2021 | Land | Fixed-bottom | - | - | 63.6 |
| Thor | 1000 | 2027 | Land | Fixed-bottom | 1977* | 58* | 53.7* |
| Nordsøen I | 3000 | >2030 | Land | Fixed-bottom | 6346 | 173 | 53.7 |
| Nordsøen II+III+Vest | 3000 - 10000 | >2030 | Energy Island(s) | Fixed-bottom | 7486 - 23318 | 129 | 61.6 |

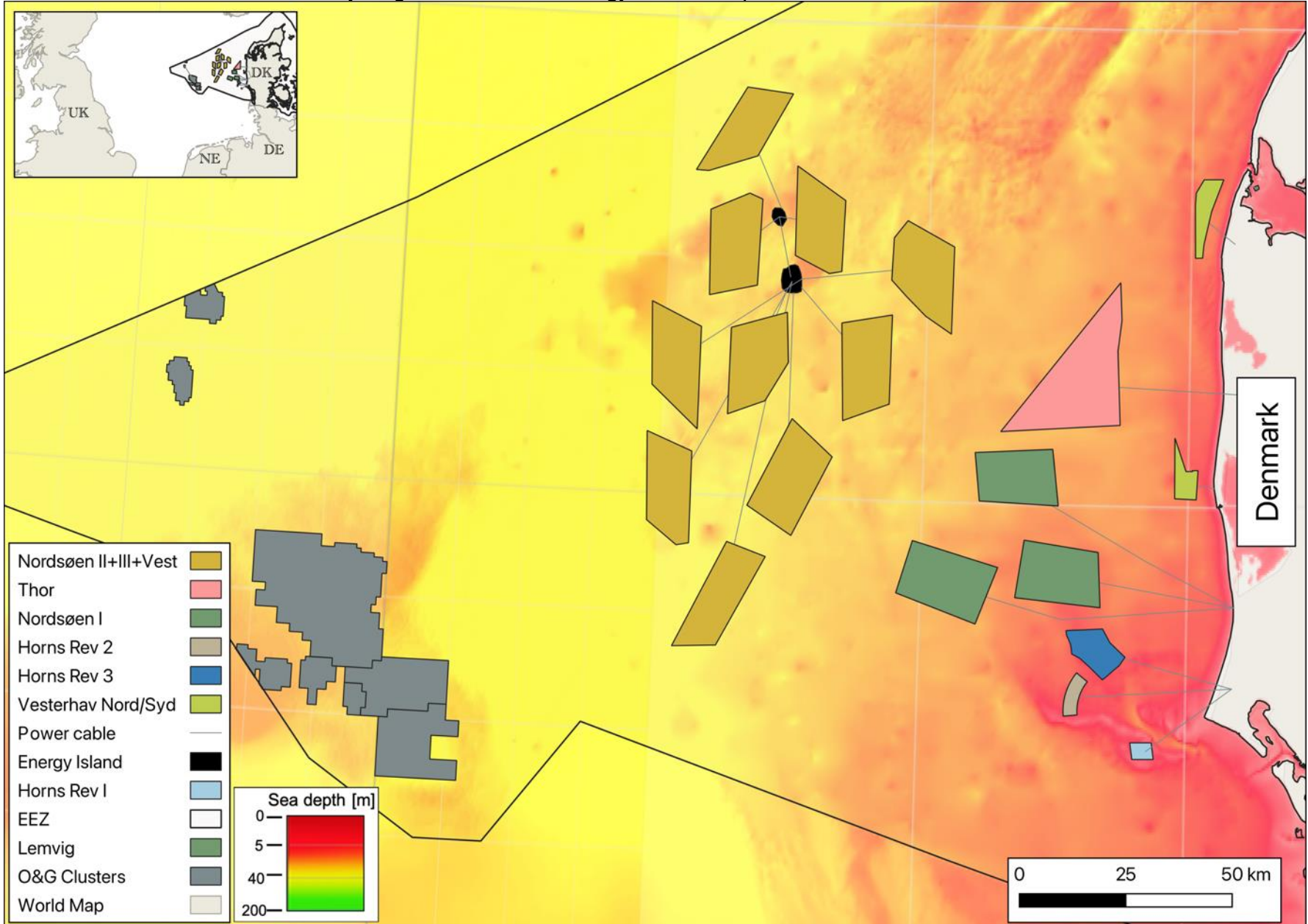

*Figure 8 Current wind farm layout in the Danish North Sea and future fixed-bottom wind tendering areas.* (4C Offshore Database 2020; COWI 2020a, 2020b)

### 3.2.2. Floating wind turbines: state of the art, costs and technology development

Floating Wind turbines (FW) are a new technology that can help harness the full potential of offshore wind in locations where the sea is too deep, or the seabed is not suitable for fixed-bottom offshore wind farms. The technology combines a regular wind turbine with a floating substructure which have been designed in multiple ways. An example of the main floating wind turbines technologies is presented in Figure 9.

In general, a floating wind turbine consists of three main parts:
- the tower, where the turbine is placed,
- the floating substructure, which supports the tower,
- the anchoring system, that fixes the substructure to the seabed with ropes or chains.

In addition, electrical cables and a substation are required to export the power to shore.

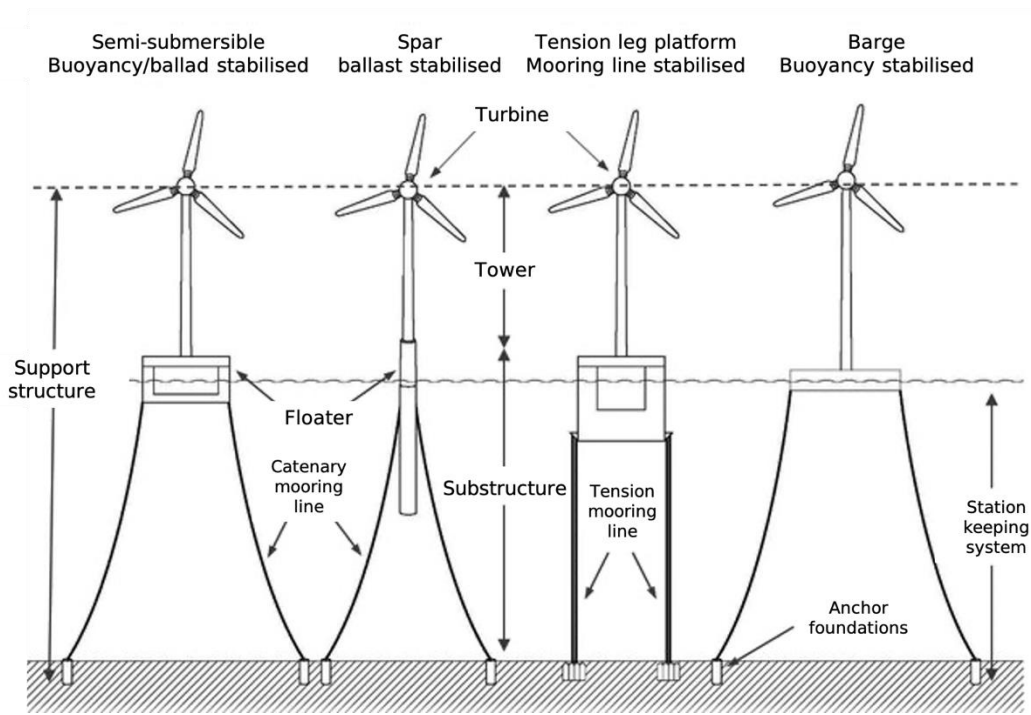

*Figure 9 Main floating wind technologies. Figure adapted from* (Afewerki 2019)

Europe's floating wind fleet is the largest worldwide (70%) with a total of 45 MW by the end of 2019 (Wind Europe 2019a). Projects up to 30 MW are already operational, while in the next few years, wind farms of 250 MW are planned to be tendered. The main markets in Europe currently include France, Spain, Portugal, Ireland and the UK. Denmark has been focused more on the research and development rather than on large scale project. This is also due to the low depth of the seabed which benefits fixed-bottom wind farms. Compared to other countries where the sea reaches great depth at relatively short distance from shore, Denmark has vast areas with low sea depth. The maximum water depth is around 50 m on the west boundary of the Danish EEZ and a distance from shore of approximately 200 km. A potential driver for the demand could be obtained from a synergic collaboration between the O&G and the offshore wind sector. The offshore platforms are located far from shore on the *Doggerbank*, which is a large sandbank with shallow water. This area is surrounded by deep waters where fixed-bottom wind farms are challenging to install. Therefore, floating wind turbines could be an optimal solution to provide renewable energy to the platforms.

Floating and Fixed-Bottom offshore wind systems have similar costs regarding the turbines and the export cable, while the installation, the foundations and the maintenance costs are specific to the foundation technology (Musial et al. 2017). A floating turbine results in high advantages regarding the installation. The turbines are assembled onshore and then towed to the offshore location. In addition, in case of large maintenance operations the turbines can be towed back to port, thus facilitating the operation and maintenance process and the decommissioning.

Currently, floating wind farms are not yet competitive with the fixed-bottom wind farms given their early stage of development. However, a decrease in costs at a similar, if not higher, pace than bottom fixed offshore wind is expected. The latter has gone from 150 €$_{2018}$/MWh in 2014 to 65 €$_{2018}$/MWh in 2017 (Wind Europe 2018).

According to (Wind Europe 2018), FW costs are expected to reach 100-80 €$_{2020}$/MWh for the first commercial scale projects using existing proven technologies and reaching final investment decision (FID) between 2023 and 2025. By this time the cumulative installed capacity would pass 1 GW in Europe. For example, France will launch a 250 MW auction in 2021 and two auctions of 250 MW each in 2022. The auctions have a target price of 120 €$_{2020}$/MWh and 110 €$_{2020}$/MWh respectively (Wind Europe 2018). Towards 2030, it is expected to reach "mature" commercial-scale and costs around 40-60 €$_{2020}$/MWh. This cost development assumes large involvement from the public and private sector. In this context, the O&G sector can support the market development with joint projects with research or commercial purposes. For example, WindFloat Atlantic is a joint project between Repsol, an O&G company and some offshore wind energy producers (Repsol 2020).

The cost assumptions used in the present study for the floating wind turbines were defined according to the aforementioned cost developments. Table 8 shows the unit costs for a given tender size for each model year towards 2050. The costs considered in this analysis are more conservative compared with the developments presented in (Wind Europe 2018), which expects costs around 40-60 €$_{2020}$/MWh in 2030. The present study considers a cost of 60 €$_{2020}$/MWh in 2030 and 40 €$_{2020}$/MWh in 2050. Moreover, the tender sizes for the floating wind farms are based on (Wind Europe 2019b) and (COWI 2020b).

Further, the floating wind turbines are assumed to be placed close to the O&G offshore platforms. The floating wind farms will provide electricity to one or more clusters of platforms. In addition, they can be interconnected to shore, meaning energy from floating wind turbines that is not consumed on the platform can be transported to shore.

*Table 8 Floating wind turbines unit costs towards 2050. (Wind Europe 2019b; COWI 2020b, 2020a) The operational costs follow the cost assumptions in Balmorel. (DEA 2021b)*

|  | Unit | 2025 | 2030 | 2035 | 2040 | 2045 | 2050 |
|---|---|---|---|---|---|---|---|
| Tender size | MW | 250 | 1000 | 1000 | 1000 | 1000 | 1000 |
| LCOE | M€$_{2020}$/MWh | 100 | 60 | 55 | 50 | 45 | 40 |
| Capex | M€$_{2020}$/MW | 3.85 | 2.35 | 2.14 | 1.96 | 1.77 | 1.58 |
| Opex fixed | k€$_{2020}$/MW/year | 41 | 27 | 25 | 24 | 23 | 22 |
| Opex variable | €$_{2020}$/MW/year | 3.0 | 3.0 | 3.0 | 2.7 | 2.5 | 2.4 |

### 3.2.3. Hydrogen production with electrolysis

This section gives an overview of technologies for offshore "green" hydrogen production based on electrolysis with renewable electricity.

### 3.2.3.1. Why hydrogen?

Hydrogen is the most abundant and simple substance in the universe. It is a colourless, odourless and tasteless element. On Earth, hydrogen is found only as part of a compound, most commonly in the form of water but also, for instance, in hydrocarbons such as methane, gasoline and coal. Thus, its use requires a separation a process which is energy intensive and the energy content of the output hydrogen is less than the energy content of the input fuel (DNV-GL 2018). Furthermore, hydrogen is generally more energy intensive to store and transport than other conventional fuels. Also, there is a high risk associated with its low ignition temperature and the mechanical degradation (hydrogen embrittlement) caused to material due to its small molecule size.

Therefore, to assess the value of hydrogen, its role in the energy transition must be considered. The primary driver for the uptake of hydrogen as an energy carrier is decarbonization. Hydrogen can be used to decarbonize the transport sector using fuel cell vehicles and hard to decarbonize sectors (e.g. cement and steelmaking industries) where hydrogen can substitute hydrocarbon fuels in the production of high-temperature heat. This implies that emphasis has to be placed on producing hydrogen in ways that allow its value chain to have a lower carbon footprint than alternative competing energy value chains, including alternative hydrogen value chains (DNV-GL 2018).

### 3.2.3.2. Hydrogen production

Hydrogen is produced with three main methods: electrolysis, reforming of natural gas and gasification of coal or biomass. Depending on the $CO_2$ emissions associated with its production, hydrogen is associated to a colour. Gas reforming, the most common method today, produces "grey" hydrogen, with 10-12 $kgCO_2/kgH_2$ (DNV-GL 2018). A more sustainable version of the "grey" hydrogen is the "blue" hydrogen. It is produced as the former, but the emissions are reduced by *Carbon Capture and Storage* (CCS) technologies. "Blue" hydrogen has a carbon footprint between 1 and 5 $kgCO_2$ per $kgH_2$. It is referred to as "green" hydrogen when it is produced by electrolysis or biomass gasification with renewable electricity and the emissions are less than 8 kg $CO_2$ per $kgH_2$ (DNV-GL 2018).

Currently, only a very small percentage of the world's $H_2$ is produced by water electrolysis because it is much more expensive than producing $H_2$ by steam Methane reforming (Esposito 2017).The single largest cost for water electrolysis is the operating expenditure (OPEX) associated with the electricity that is used to drive this endothermic reaction. However, with the increase in renewables energy capacity, such as wind energy, in particular in the North Sea, there is a large availability of cheap energy that could drive the costs down.

Another limitation to use of hydrogen is its transportation. Pipeline transport of compressed gaseous hydrogen is in general the most cost-effective way of transporting large volumes over long distances (DNV-GL 2018). However, due to the size of its molecules, it can leak and enhance fractures in common steel pipelines, therefore existing gas pipelines must be upgraded or a new dedicated hydrogen pipelines must be used.

An alternative solution is blending hydrogen into natural gas up to a certain volumetric percentage, around 10% (Snam 2020). This approach enables the use of existing gas pipelines resulting in large cost savings. However, many European countries don't allow this mixture in the gas distribution network, therefore a separation station is required to extract hydrogen at its destination, thus transportation costs increase.

### 3.2.3.3. The Hydrogen market

Besides the aforementioned reasons that make hydrogen a promising yet challenging energy fuel and carrier, there is already a large and thriving industry relying on it. According to the Hydrogen Council, the world currently consumes more than 55 Mt per year, of which around 95% originate from fossil fuels. The majority of this $H_2$ is used for ammonia production (55%), in petroleum refining (25%), and for methanol production (10%). The energy required to produce 55 Mtpa $H_2$ represents about 3% of the global energy demand (DNV-GL 2018).

In Europe, hydrogen has a central role in the European Green Deal that aims at a climate-neutral EU in 2050.  (European Commission 2020) reports that the share of hydrogen in Europe's energy mix is projected to grow from the current less than 2% to 13-14% by 2050.  According to the *European Clean Hydrogen Alliance*, the electrolysis industry requires a strong ramp-up. Renewable hydrogen electrolysers in EU are expected to reach a total capacity of 6 GW by 2024 and 40 GW by 2030. The hydrogen demand, according to the 2x40GW Green Hydrogen initiative, will be 16.9 Mt or 665 TWh (Hydrogen Europe 2020). Moreover, the study assumes 7.4 Mt of hydrogen to be supplied by green hydrogen, 4.4 Mt is produced in the EU, while 3 Mt is imported from North Africa and Ukraine.

As part of the Green Deal, Denmark is committed to the development of the renewable hydrogen sector through large investments. The Danish Ministry of Climate, Energy and Utilities supports two project, "GreenLab Slice PtX" and "HySynergy", for a total of 17 million Euros (State of Green 2019). Further, the Danish government supported the HRES project with around 4,6 million euros . HRES is a demonstration project using offshore wind power to produce renewable hydrogen for road transport. It consists of a 2 MW electrolysis plant producing hydrogen from renewable electricity provided by two 3.6 MW wind turbines (Ørsted 2019). The project is led by Ørsted A/S together with 7 industrial partners.

### 3.2.3.4. The electrolyser technology outlook

In general, three main electrolysis technologies are available to produce hydrogen from water: AEL (Alkaline Electrolyser), PEMEL (Polymeric Electrolyte Membrane Electrolyser) and SOEC (Solid Oxide Electrolyser Cell) (VTT 2018) . AEL is the most established technology with relatively low capital costs, but high maintenance costs due to the fact that the electrolyte is highly corrosive, and circulation of the electrolyte is required. PEMEL systems provide higher current densities but have shorter lifetimes and higher costs. The SOEC systems are the least developed but are expected to have improved efficiency especially if the integration of waste heat is possible.

These electrolyser technologies consist of electrolyser cells that are combined to build an electrolyser stack. To build a Gigawatt scale electrolyser, a number of electrolyser stacks are placed in parallel. These electrolyser technologies are expected to achieve remarkable technology improvements in the next decade (Hydrogen Europe 2020). Amongst other things, these include higher efficiencies, less degradation, higher availability, larger cell sizes, higher operating pressure, less critical material use together with overall reduced material use, which will reduce hydrogen production cost by electrolysers (Hydrogen Europe 2020).

Moreover, next to these technology improvements, economies of scale – higher installed capacity volume and plant size – will bring down the electrolyser cost. Therefore, on the one hand, automated production of the electrolyser cell components, cells and stacks will bring down the cost for the electrolyser stacks. On the other hand, building Gigawatt scale electrolyser plants will reduce the balance of plant costs per kW. The balance of plant costs are the costs for compressors, gas cleaning, demineralised water production, transformers and the installation cost. A substantial electrolyser market volume together with realizing GW scale electrolysers, are essential drivers for significant cost reductions (IEA 2020).

Table 9 provides an overview of the techno-economic data for the electrolysers from two reports (IEA, Hydrogen Europe) and a real-life case study (DNV-GL). One can notice that the current costs of electrolysers and their development towards 2050 have a wide range. Hydrogen Europe assumes lower costs than the IEA throughout the whole period. The cost reduction between 2020 and 2050 is between 50% and 60% in both sources. Moreover, there is a significant difference in the LCOE in 2050 between IEA and Hydrogen Europe. This is due to the large number of variables that are involved in its calculation. Among other things, a low electricity price is a key element to reach a low LCOE. According to Hydrogen Europe, it accounts for 60-80% of the hydrogen cost – every 10 €/MWh less electricity cost at 80% electrolyser efficiency HHV (Higher Heating Value), translates to a hydrogen cost reduction of 0.5 €/kg (Hydrogen Europe 2020)

In the present study, the hydrogen plants is based on the techno-economic data provided in the case study of DNV-GL (DNV-GL 2018). It plans to install an electrolyser plant of 200 MW on two platforms in the Dutch North Sea in 2025. The electricity is provided by wind turbines and the hydrogen produced is transported via pipelines to shore. Therefore, there are strong similarities between the presented case study and the analysis performed in the present study.

*Table 9 Techno-economic data for hydrogen production by electrolysers. All costs in €$_{2020}$. *The cost development of DNV-GL is based on the IEA's projections. LHV: Lower Heating Value. HHV: Higher Heating Value*

| Source | Parameter | Units | 2020 | 2030 | Long Term |
|---|---|---|---|---|---|
| **IEA** | Capex | €/kW | ~800 | ~620 | ~400 |
| | Opex | %/year Capex | 1.5 | 1.5 | 1.5 |
| | Efficiency (LHV) | % | 64 | 69 | 74 |
| | LCOE | €/kg H$_2$ | 2.6 – 6.4 | - | 1 – 2.7 |
| | El. Price | €/MWh | - | - | ~40 |
| **Hydrogen Europe** | Capex | €/kW | 300-600 | 250-500 | <200 |
| | Opex | %/year Capex | 1.5 | 1 | <1% |
| | Efficiency (HHV) | % | 75-80 | 80-82 | >82% |
| | LCOE | €/kg H$_2$ | 1.5-3.0 | 1.0-2.0 | 0.7-1.5 |
| | El. Price | €/MWh | 25-50 | 15-30 | 10-30 |
| **DNV-GL** | Capex | €/kW | 1170 | - | 400* |
| | Opex | %/year Capex | 2 | - | 2* |
| | Efficiency (LHV) | % | 67 | - | 74* |

The cost assumptions provided in the case study are projected to 2050 based on the cost developments estimated by the IEA (see above). The resulting techno-economic assumptions are provided in detail in Table 14 in the appendix. A base unit of 200 MW plants is considered, which can be stacked to reach higher capacities. Accordingly, the costs will increase linearly with the increase in capacity. Hydrogen produced from the platform is transported to shore by new dedicated pipelines (36" Ø) at a cost of 179 k€/GW/km (Global CCS institute 2020). Alternatively, it is assumed that existing gas pipelines can be used for 10% costs of new pipelines.

### 3.2.4. Carbon Capture and Storage (CCS)

This section gives an overview of the CCS technology, in terms of state of the art (section 3.2.4.1), application in depleted O&G fields (3.2.4.2) and techno-economic assumptions for this study (3.2.4.3).

### 3.2.4.1. CCS state of the art

Carbon capture and storage (CCS) is the process of capturing $CO_2$, compressing it for transportation and then injecting it deep into a rock formation at a carefully selected and safe site, where it is permanently stored. The three main steps are:

- **Capture:** the separation of $CO_2$ from other gases produced at large industrial process facilities such as coal and natural-gas-fired power plants, steel mills, cement plants and refineries.
- **Transport:** once separated, the $CO_2$ is compressed and transported via pipelines, trucks, ships or other methods to a suitable site for geological storage.
- **Storage:** $CO_2$ is injected into deep underground rock formations.

CCS is considered a key technology to reduce $CO_2$ emissions in sectors which are hard to decarbonize such as the manufacture of industrial materials like cement, iron and steel and pulp and paper. Carbon separation/capture technologies have been operational on a large-scale in the natural gas and fertiliser industries for decades and have recently become operational in the power sector (Global CCS institute 2020).

 Today, there are 65 commercial CCS facilities in the world, of which 26 are operating, 3 under construction, 34 under development and 2 have suspended operations (Global CCS institute 2020). There are another 34 pilot and demonstration-scale CCS facilities in operation or development and eight CCS technology test centres CCS facilities currently in operation can capture and permanently store around 40 Mt of $CO_2$ every year. Overall, it is considered that the global geological storage capacity for $CO_2$ is many times larger than what is required for CCS to support fully in the achievement of net-zero emissions under any scenario (Global CCS institute 2020).

### 3.2.4.2. CCS in depleted O&G fields

Globally, there are a number of examples of where $CO_2$ storage in depleted or depleting O&G fields has been proven (a complete overview is provided in (Global CCS institute 2020)). There are many advantages from the use of depleted O&G reservoirs. First all, they have already demonstrated their capacity to store gas and other fluids for a very long time. Moreover, there is a great knowledge about the geological properties of the reservoirs due to the activities of O&G production and exploration. Specifically, through the exploration it is possible to estimate the physical structure of rock formations, to locate where the best potential $CO_2$ storage sites may be and to offer insights into the fluid flow characteristics of the rock formation (Lai et al.). Furthermore, information about the production of oil and gas, namely the total amount produced, and the production rates can provide a good first estimate of the amount of $CO_2$ that can be stored. Finally, the surface facilities are very useful and relatively easy to convert for the use of $CO_2$ storage (Ajayi et al. 2019).

It is important to mention that not all O&G reservoirs are suitable for carbon capture and storage. According to (Höller and Viebahn), the Danish O&G fields provide only minor capacities with 0.2 Gt. Most Danish hydrocarbon reservoirs are composed of chalk carbonate characterized by very low matrix permeabilities, which may cause concern that the injectivity is prohibitively low (Bech and Frykman 2003). Sandstone was proven to be more suitable. Among the 10 CCS projects presented in (Hannis et al.), 7 are realized in a sandstone reservoir. Few Danish fields are sandstone, namely Harald West and Lulita, and Siri, Cecilia, and Nini, that are operated by the Total E&P Danmark A/S together with the DUC and INEOS Oil & Gas Denmark, respectively. The latter fields have been considered for a CCS project called GreenSand. A description of the project is provided in section 2.4.3.2.

### 3.2.4.3. Techno-economic assumptions for CCS

A high level techno-economic analysis was performed on the Harald West field to gain an approximate estimate of the potential for storing $CO_2$ in its reservoirs. The cumulative gas production of the Harald field is 24 $BNm^3$ , however the specific production of Harald West is not known (DEA 2021a). The storage efficiency factors to globally estimate the storage capacity in depleted fields is around 75% (Global CCS institute 2020). Therefore, about 18 Mtons of $CO_2$ could be stored in the Harald Field.

The costs of capturing and storing $CO_2$ extracted from natural gas is in the order of 12-20 $€_{2020}$ /ton, where the low cost is applied when the plant is close to the $CO_2$ injection site and the upper end of the range includes some transportation (Global CCS institute 2020). Moreover, the low costs are due to the availability of high purity $CO_2$ which typically requires only dehydration before it can be compressed and stored. The profit from storing $CO_2$ are calculated as the difference between the $CO_2$ tax (see section 4) and the cost of capturing and storing it.

# 4. Energy system modelling with Balmorel

The second part of this analysis involves a whole system analysis with the Balmorel model, as presented in this section. Section 4.1 first gives an introduction to the Balmorel model, before section 4.2 defines the scenarios employed in this analysis. Section 4.3 then presents and analyses the results according the typology of scenarios from section 4.2, culminating in the explorative roadmap for the sector as well as a sensitivity analysis. Section 4.4 then discusses both the employed method and the results in a broader context.

## 4.1. Introduction to the Balmorel model

Balmorel is an open source, deterministic, technology-rich, bottom-up, partial equilibrium optimisation model, as described in (Wiese et al. 2018). It minimises the total costs of the system, including investments, variable and fixed costs components, whilst being constrained by technical, physical, or regulatory aspects. It is designed to simultaneously optimise the generation, transmission, and delivery to end-use sectors of heat and electricity for an energy system. Inputs include the final energy demand, which must be satisfied in each time step, as well as fuel prices and the availability of different resources like wind and solar energy. In the context of this study, the model optimises the energy system from a socioeconomic point of view, meaning it adopts the perspective on one central decision maker (central planner).

Electricity and heat generation and storage units are described by their technical and economic characteristics. Each technology has specific initial investment, variable and/or fixed fuel and/or operation & maintenance costs (O&M), efficiency, technical lifetime and used fuel type. The existing planned and installed capacity of each generation technology is defined exogenously, together with its commissioning and decommissioning date. However, it is possible to allow the model to install or decommission extra generation capacity endogenously, as a result of the optimisation.

The flexibility regarding temporal and spatial resolution allows for a wide range of modelling options in Balmorel. Time steps are organised by three hierarchical layers from higher to lower: Years (Y), Seasons (S), and Terms (T). The years considered during this study are: 2025, 2030, 2035, 2040, 2045, and 2050. Each year is divided in 16 seasons (S) which represent weeks, and 24 terms (T) which are intended to represent each week as 3 consecutive days in which only 1 hour every 3 is considered. The level of foresight of the model is flexible and it depends on the settings decided by the user. In this case, two years are considered at the same time, which means that both 2035 and 2030 are considered when optimising 2030, for example. This is known as a rolling horizon approach.

Regarding the spatial resolution, Balmorel is also divided in three hierarchical layers from higher to lower: Countries (C), Regions (R), and Areas (A). Each country can be divided into one or multiple regions, and each region can be divided at the same time in one or multiple areas. Regulatory targets and policy measures can be defined by country, such as the minimum renewable energy required or a $CO_2$ emissions limit. The countries included in the optimisation are presented in Figure 10.

Electricity demand is defined by regions, in which the electricity balance is maintained. Areas are mainly used to define heat demand and climate conditions such as wind, solar, and hydro potential for renewable energy generation. Furthermore, the type of technologies that can be installed in each area are defined as an input. For example, the same region can have three different areas defining onshore, nearshore, and far offshore wind conditions respectively. Curtailment of variable renewable energy generation is allowed if the model finds it optimal, i.e. if it is cheaper than the alternatives, such as investing in storage, networks, Demand Side Management (DSM) etc.

During this study, the total installed capacities for both electricity and heat generation, storage, and transmission, are cost-optimised for each simulated year. The partly known and expected developments in offshore wind in the Danish North Sea are modelled based on published data in (4C Offshore Database 2020; COWI 2020a, 2020b). Finding the optimal mix of total installed capacity for each area of the model demands significant computational time. Hence, to reduce the model's complexity a low time resolution, as explained above, is employed in this study.

In order to model the existing Oil & Gas platforms, the clusters of platforms – Gorm, Tyra, Halfdan, Dan, and Harald – are defined as Regions in the model. Each region is assumed to have a natural gas turbine installed and a constant annual electricity demand that needs to be satisfied. The techno-economic assumptions of the clusters are provided in section 3.1.

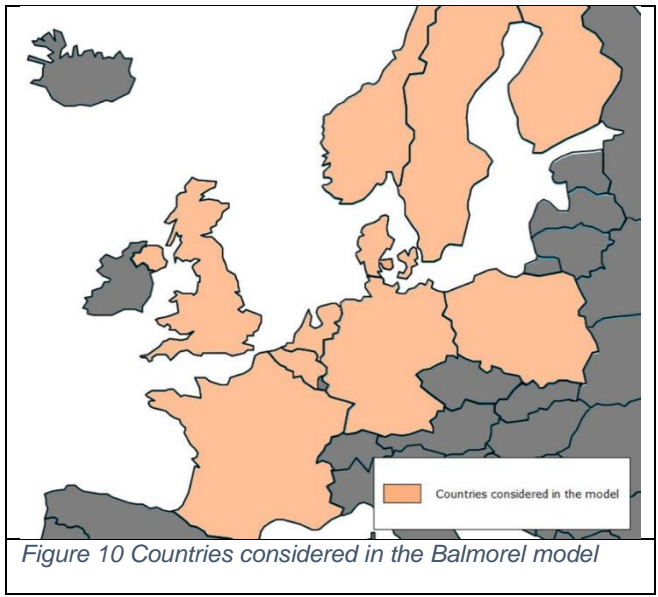

*Figure 10 Countries considered in the Balmorel model*

The decommissioning cost-savings from re-purposing the O&G platforms are modelled as a technology with a negative investment equal to the decommissioning costs. After the decommissioning date of each platform, the model can invest in this technology in the platform's regions, reducing the total system costs while considering the operational and maintenance costs of the platform.

It is expected that hydrogen will have a significant impact in the European energy system in the future. Hydrogen is expected to be crucial for the production of fuels that allow the decarbonisation of the transport sector alongside electric vehicles. Bio- & electro-fuels could account for a significant percentage of the total electricity demand in the future (S&P Global 2020). To account for this fact in this study, a significant additional electricity demand is considered related to the decarbonisation of the transport sector. This electricity demand is modelled in a simplistic way by two constraints, to avoid increasing substantially the complexity of the optimisation problem. The first constraint ensures that the annual energy required to decarbonise the transport sector is produced. The second constraint restricts when and where this extra demand can be covered, by limiting the ratio between the additional transport demand, and the annual average demand of each region (excluding transport). Referring to Figure 10 above, this additional electricity demand for hydrogen can be met in any or all of the countries within the model's scope.

The assumed price developments for natural gas and $CO_2$ are shown in Figure 11 below, which represents typical expected future scenarios based on previous work and Danish Energy Agency projections. The natural gas price increase from 20 $€_{2012}$/GJ in 2020 to 39 $€_{2012}$/GJ in 2050. The gas price used in the gas turbines on the platforms is considered as an opportunity cost, i.e. the forgone cost of selling this gas. For the $CO_2$ tax, the *E&R* curve shows the default trend assumed in the *E&R* scenario (see section 4.2), which rises to about 80 $€/tCO_2$ and 130 $€/tCO_2$ in 2030 and 2050 respectively, and reflects the progressive tightening of $CO_2$ allowances in the EU ETS.

In addition, two alternative scenarios CO2-mod and CO2-low assume a linear development to 80 €/tCO2 and 60 €/tCO2 in 2050 respectively.

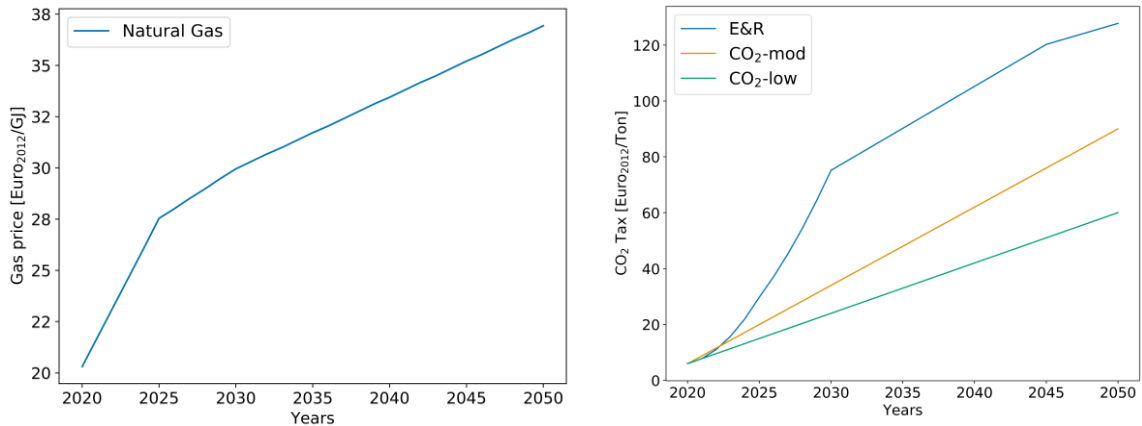

*Figure 11 Natural gas and CO2 tax evolution through the years considered in the model.* (DEA 2021b)

## 4.2. Scenario analysis

This section gives an overview of the analysed scenarios employed in this analysis. The Business as Usual (BAU) scenario (section 4.2.1) reflects the decommissioning of all O&G platforms based on the schedule outline in section 3.1.4. The Electrification and Repurposing (E&R) scenario (section 4.2.2) represents a future in which the Balmorel model is able to electrify the platforms and/or repurpose them once they reach the end of their useful life. In addition, several sensitivity scenarios (section 4.3.4) are intended to explore the impact of key assumptions on the results.

### 4.2.1. Decommissioning (Business as Usual, BAU)

The *BAU* scenario is based on the following central assumptions (Figure 12):

- Current (2021) operations of the platforms continue.
- The current conditions on the platforms are maintained until the time of decommissioning.
- Platforms are considered in the context of clusters.
- No investments are carried out on the platforms.
- The platform's clusters are decommissioned according to the decommissioning timeline (cf. section 3.1.4).
- The decommissioning process consists in shutting down all operations and removing the infrastructure.
- The platforms' main power (and heat) supply is from gas turbines at all times.

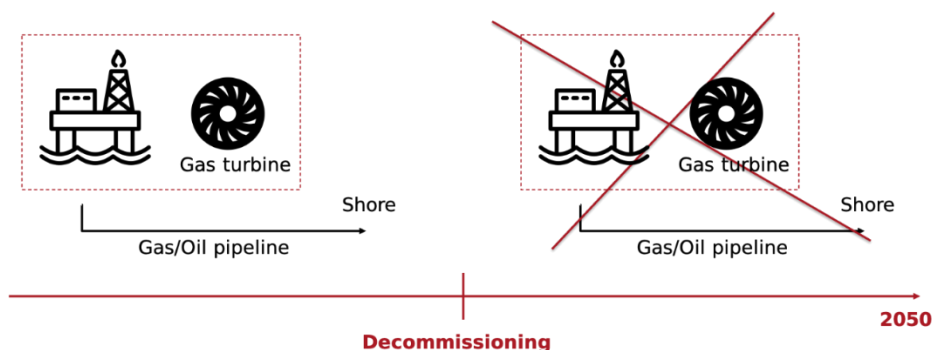

*Figure 12 Schematic of the Decommissioning / BAU scenario*

### 4.2.2. Electrification & Repurposing (E&R) Scenario

The Electrification & Repurposing (E&R) scenario is the main scenario, in which new hydrogen pipelines are allowed, floating wind turbines, interconnection with transmission lines (Figure 13). It is based on the following assumptions:

- The model can invest in several technologies on the platform to reduce the total system cost.
- The platform can be powered from wind farms or from the electricity grid to the expense of the gas turbines, which are decommissioned.
- As an alternative to decommissioning, the platforms can be repurposed. Some of the existing infrastructure can be reused. For example, the pipelines could transport hydrogen or $CO_2$.
- The repurposing of the platform is assumed to be possible only after the production has ceased.
- The repurposing of the platforms relies on the removal and replacement of the topside, based on costs and assumptions set out in 3.1.7.

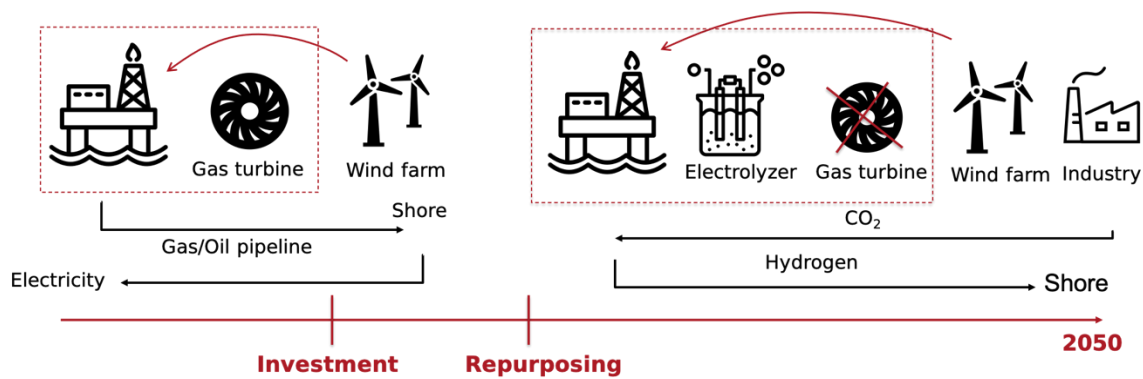

*Figure 13 Schematic of the Electrification & Repurposing (E&R) scenario*

## 4.3. Results analysis

This section presents and analyses the results from the Balmorel model. It is broadly structured according to the scenarios introduced in the previous section.

### 4.3.1. Decommissioning (BAU)

In the decommissioning scenario, the clusters of platforms follow the current operational settings until decommissioning (cf. Figure 14). In Table 15, the cluster's annual energy related expenses are provided. $CO_2$- and Fuel-related expenses represent the highest share of costs among all clusters. On average, they account for 17% and 79% of annual expenses in 2025, respectively.

Furthermore, an increase in yearly costs for all platforms towards 2050 can be noticed. The reason behind this growth is related to the upward trend of the costs of natural gas and $CO_2$ emissions. Among all clusters, the yearly average growth is at least 9%; the peak is observed in 2030 with an average growth of 33%. In Figure 11, one can notice that between 2025 and 2030 the $CO_2$ tax has the highest increase which explains the peak here. In addition, the shares of yearly costs changes overtime. $CO_2$ related costs almost double between 2025 and 2030 among all cluster at the expenses of the Fuel costs.

As shown in Figure 14 below, the *BAU* scenario involves the full decommissioning of each cluster/field according to the schedule outlined in section 3.1.4. The years without any costs thereby represent years in which the platform is no longer operational. The operational costs of the platform, namely the expenses related to drilling and processing the hydrocarbons fuels, ranges from 74% to 144% of the cumulative energy related yearly costs among the different clusters. In addition to the total energy-related OPEX of 736 M€$_{2012}$ for all clusters, not shown in Figure 14 are the decommissioning costs of 6,352 M€$_{2012}$ and platform OPEX (e.g. drilling, oil & gas processing) of 185 M€$_{2012}$.

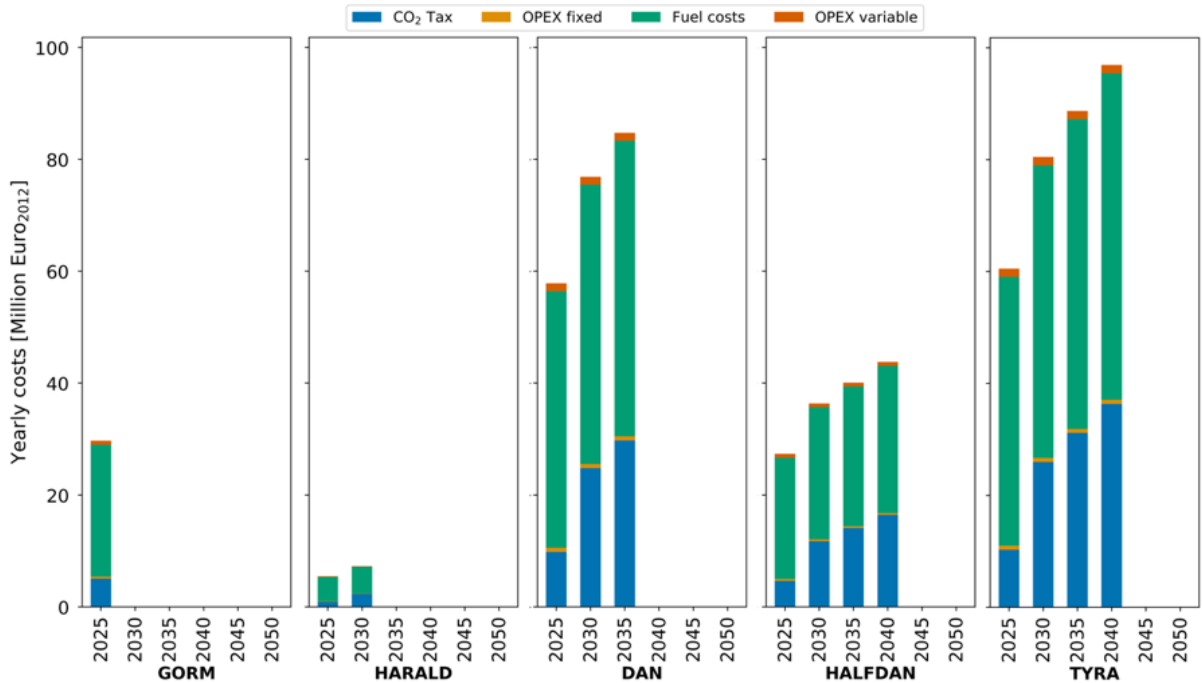

*Figure 14 Yearly energy related expenses breakdown for each cluster towards 2050. The figure does not take into account the operational costs of the producing the oil & gas (e.g. drilling, oil & gas processing).*

### 4.3.2. Electrification & Repurposing (E&R)

In the first modeling year of 2025, all clusters are electrified through connections to the shore or the energy island (rather than to offshore wind plants, for example) and the gas turbines are decommissioned. In Figure 18, these interconnections are presented with a blue solid line.

The platform electrification results in large savings in terms of Costs and $CO_2$ emissions as reported in Table 16. On average, the reduction in Costs and $CO_2$ emissions through electrification is 72% and 85%, respectively. This results in aggregated savings of 129 M€$_{2012}$ and 1 Mt of $CO_2$ emissions in 2025. Due to electrification, the electricity price on the platforms decreases from 160 to 43 €$_{2012}$/MWh (in the case of the gas turbines the former represents a shadow price). The savings in $CO_2$ emissions can contribute strongly to reach Denmark's objective of reducing 1.4 Mtons of greenhouse gas emissions per year by 2025 (S&P Global 2020). Moreover, due to the electrification, the O&G sector accounts for just 2% of Denmark's total $CO_2$ emissions in 2025, compared to around 15% in the *BAU* scenario. The aggregated costs across the cluster in 2025 for the *E&R* scenario are 57 M€$_{2012}$ (excluding operational costs of the platform) and the $CO_2$ emissions are 176 ktons.

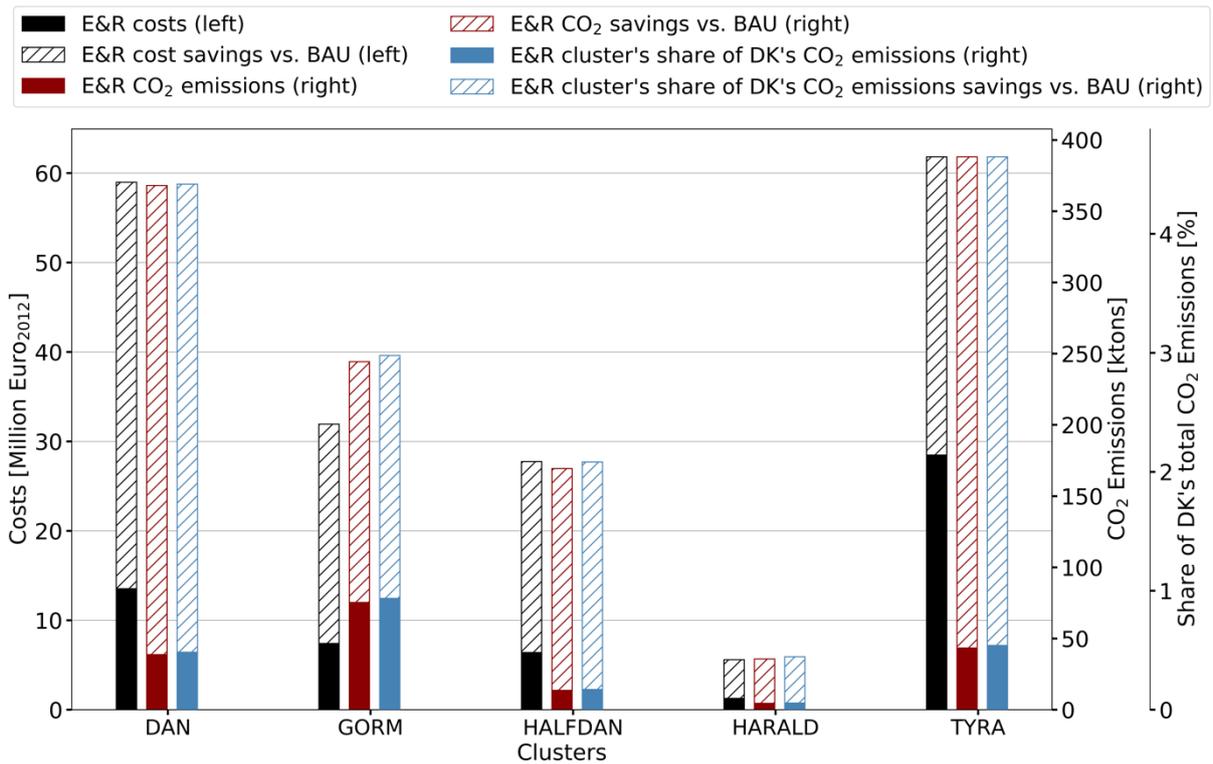

*Figure 15 Cost and $CO_2$ emissions savings in 2025 for each cluster of platforms. The left-hand side y-axis shows the Costs while on the right-hand side the y-axes show the $CO_2$ emissions and the share of the cluster's emission in Denmark's total $CO_2$ emissions.*

As an alternative to decommissioning, the clusters of platforms are used for hydrogen production, carbon capture and storage, and export of electricity from floating wind farms. In Figure 16 the average breakdown of the cluster's yearly costs is shown.

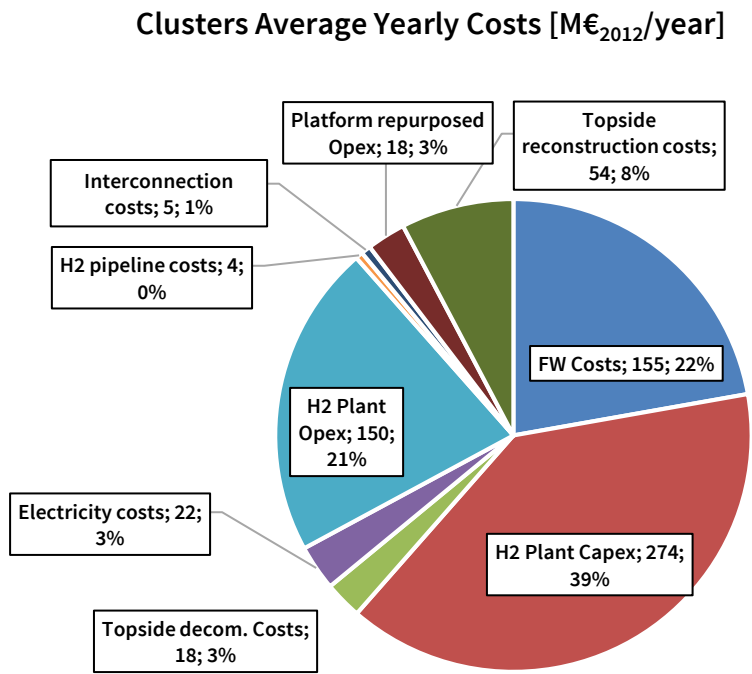

*Figure 16 Clusters' average annual costs in E&R scenario*

One can notice that the largest share are related to the hydrogen plant and the floating wind farm, with 60% and 22%, respectively. In absolute terms, the average yearly expenses are 428 M€[2012]/year and 155 M€[2012]/year, respectively. The remaining part of the costs consists in repurposing and operating the platforms; it accounts for 14% (90 M€[2012]/year).

In Table 10 the floating wind farms and the electrolyser plants capacities in 2050 are presented for each cluster. The cumulative capacities are 5.8 GW and 3.6 GWel, respectively, whereby the ratio of electrolyser to wind capacity is capped at a maximum of 80%. From the distribution of the capacity among the clusters, one can notice a correlation between the two technologies. The availability of cheap electricity from floating wind is a driver for the model to allocate electricity demand for the decarbonization of the transport sector on the platforms. The hydrogen LCOE in 2050 is 3.6-6.8 €[2012]/kg $H_2$ (average 4.89 €[2020]/kg $H_2$), which is higher than the expected price of renewable hydrogen of 0.7-1.5 €/kg $H_2$ already in 2030 (Reuters 2020). However, the hydrogen plant costs, and the expenses related to repurposing the platform are highly uncertain. Therefore, the LCOE alone does not indicate the feasibility of the hydrogen production offshore. The aggregated hydrogen production is about 505 kt of $H_2$, which is far below the hydrogen demand of 4.4 million tons assumed by the 2x40 GW Green Hydrogen initiative (Ad van Wijk and Jorgo Chatzimarkakis 2020) to be produced by green hydrogen in 2030.

*Table 10 Installed capacities of floating wind and electrolyser in 2050 for all clusters*

| Technology | Unit | Dan | Gorm | Halfdan | Harald | Tyra | Total |
|---|---|---|---|---|---|---|---|
| Floating wind | GWel | 1.7 | 0.9 | 1.0 | 0.2 | 2.0 | 5.8 |
| | % | 29% | 16% | 17% | 3% | 34% | 100% |
| Electrolyser | GWel | 1.3 | 0.5 | 0.3 | 0.1 | 1.3 | 3.6 |
| | % | 37% | 15% | 8% | 3% | 37% | 100% |

An alternative use of the platforms and the wells is Carbon Capture and Storage (CCS) as discussed in section 3.2.4. The value from the implementation of this technology lies in costs of emitting $CO_2$ that are saved. In Table 11, the unitary profit on $CO_2$ storage are presented for the *BAU* scenario and the sensitivity scenarios *$CO_2$-mod* and *$CO_2$-low*. The latter scenarios include a lower $CO_2$ tax compared to the *BAU* scenario as shown in Figure 11. Among all scenarios, from 2030 CCS yields a positive return, while, in 2025 only the *BAU* scenario is profitable.

Given these profits and the storage capacity assumed in Harald of about 18 TNm³, we can assess a value of the reservoirs of between 59 and 981 T€[2012] in 2030 and 708 and 1928 T€[2012] in 2050 according to the three scenarios.

*Table 11 Unit profit [€[2012]/tCO₂] of CCS with CO₂ tax in Figure 11. CCS costs are about 17 €[2012]/tCO₂.*

| Scenario | 2025 | 2030 | 2035 | 2040 | 2045 | 2050 |
|---|---|---|---|---|---|---|
| BAU | 9 | 54 | 70 | 85 | 100 | 107 |
| $CO_2$-mod | -1 | 13 | 27 | 41 | 55 | 69 |
| $CO_2$-low | -6 | 3 | 12 | 21 | 30 | 39 |

### 4.3.3. Roadmap

The performed analysis has highlighted a possible roadmap for the Danish Oil & Gas sector towards 2050 as presented in Figure 17. The clusters of platforms cease operation according to the timeline provided in Figure 17, while Balmorel, the optimization model, can choose to invest on the platforms (e.g. floating wind farms, electrification) and decommission existing infrastructure on the platform (e.g. gas turbines). It is important to emphasize that the model optimizes only two adjacent years, i.e. the current year and the future one (e.g. 2025 and 2030). Therefore, some choices can result not optimal if analyzed in the time range from 2025 to 2050.

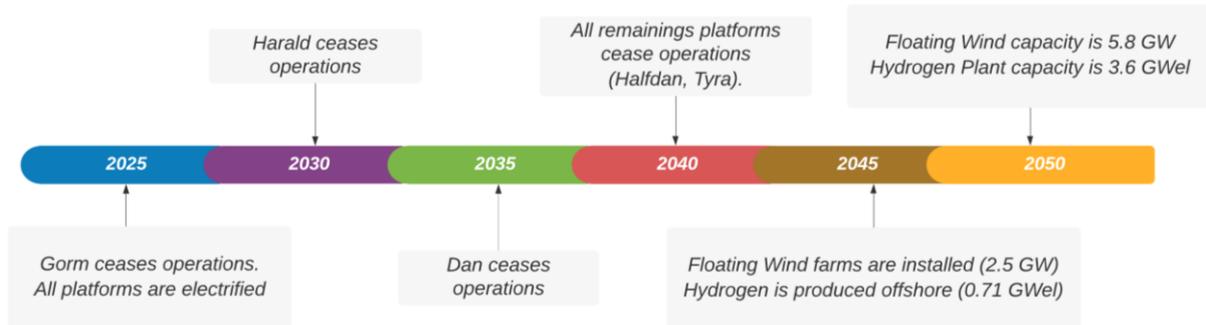

*Figure 17 Timeline of the Electrification and Repurposing (E&R) scenario.*

At first, in 2025, all platforms are electrified through a connection from shore or from the energy island which can be interconnected to other countries. In Figure 18, these interconnections are graphically represented by a solid blue line. Between 2025 and 2035 the layout is unchanged, there aren't new interconnections. In 2040, the platforms are interconnected within each other and a secondary cable is connected to the energy island which is now also connected to shore. In 2045, Harald, Halfdan and Tyra invest in floating wind farms; the cumulative capacity is 2.5 GW. In addition, there is hydrogen production on all platforms except for Gorm. The aggregated capacity is 0.71 GWel, where Tyra and Halfdan account for 98% of it. The hydrogen produced is transported to shore by pipelines, which are connected to all producing platforms. Furthermore, on this year, the energy island has 10 GW of installed wind capacity therefore there is a large availability of green energy offshore which benefits the production of hydrogen offshore.

Finally, in 2050, the layout changes slightly from 2045. In terms of interconnections, there is no change but an increase in capacity. Floating wind capacity doubles and reaches 5.8 GW. Hydrogen is produced on all platforms and the aggregated electrolyser capacity increases to 3.56 GW, more than 5 times compared to 2045. Overall, it can be noticed that the energy island works as a bridge between the platforms and the Danish shore which are interconnected with a 1.6 GW power cable.

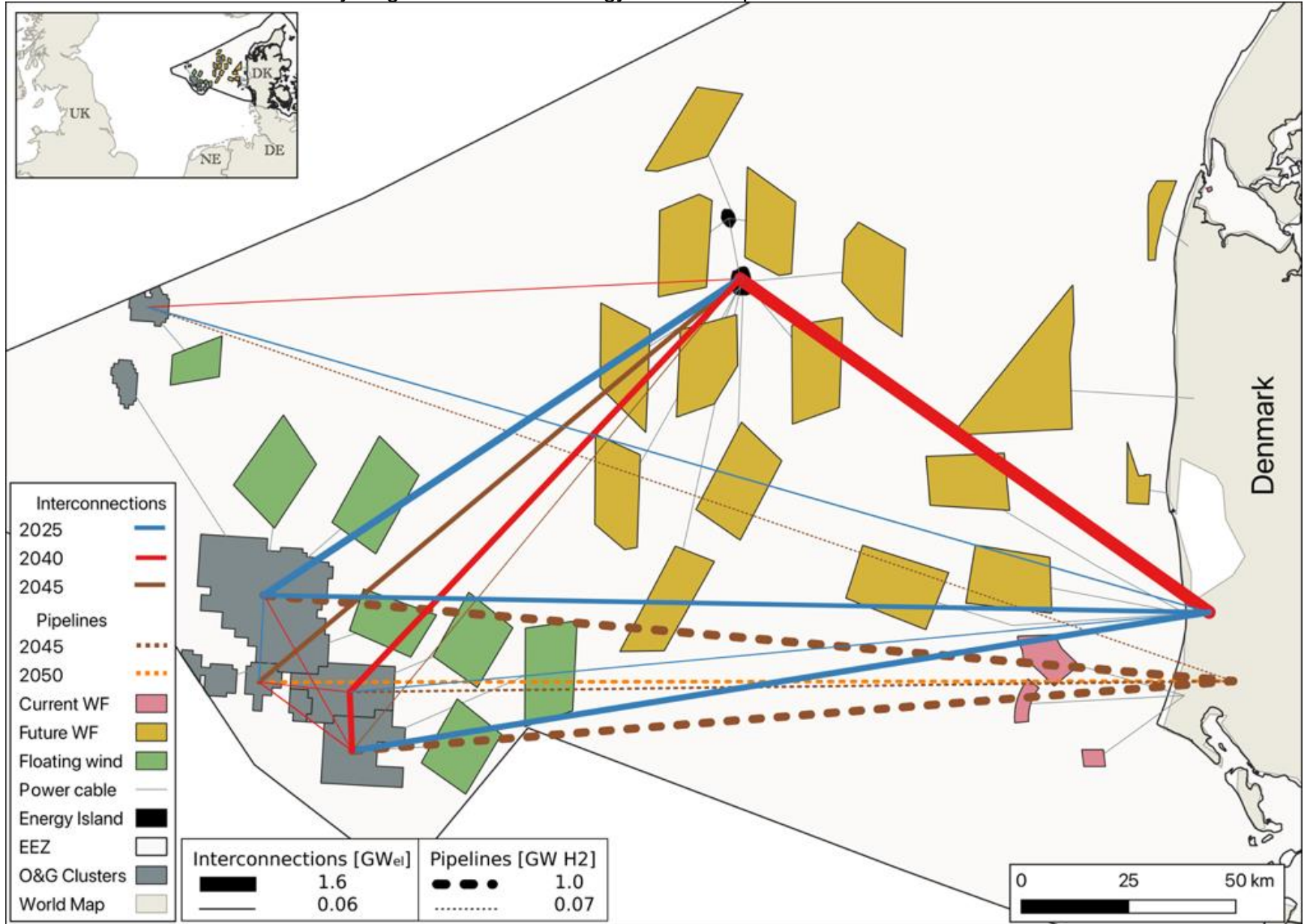

*Figure 18 Overview of Danish offshore region with developed energy system infrastructure to 2050 in E&R scenario*

### 4.3.4. Sensitivity analysis

The sensitivity analysis includes seven additional scenarios, which explore the impact of uncertain input data on the results. They are based on the *E&R* scenario and analyse the key factors for driving the total system costs. They include the following (Table 12):

- The development in the floating wind costs are uncertain, hence the *FW-high* and *FW-low* scenarios account for this.
- In the *E&R* scenario, the model only has the option to invest in new $H_2$ pipelines, whereas *$H_2$-low* reflects the assumption of reduced costs (10% of a new pipeline) to reuse existing oil and/or gas pipelines.
- The assumed *$CO_2$ Tax* development is reduced in two more moderate scenarios *$CO_2$-low* and *$CO_2$-mod*.

*Table 12 Sensitivity analysis scenarios based on the E&R scenario*

| Scenario | Element | Unit | Variation |
|----------|---------|------|-----------|
| FW-high | Floating Wind turbines LCOE | €/MWh | +25% |
| FW-low | Floating Wind turbines LCOE | €/MWh | -25% |
| TL-25, TL-50 | Electricity transmission line | €/MWh | +25%, +50% |
| $CO_2$-low | $CO_2$ Tax | €/tCO | Linear increase from 8 to 60 €/tCO in 2050. |
| $CO_2$-mod | $CO_2$ Tax | €/tCO | Linear increase from 8 to 90 €/tCO in 2050. |
| $H_2$-low | Reuse of existing gas pipeline to transport hydrogen | €/MW/km | $H_2$ pipelines costs 10% of a new pipeline. |

### 4.3.4.1. LCOEs of electricity on the platform

This analysis focuses on the development of the costs of powering the platforms from multiple energy sources. It evaluates the levelized cost of electricity (LCOE) that result from powering the platform by different energy sources: the gas turbine and a floating wind farm or through an interconnection to shore. For each of these sources, the LCOE has been calculated based on three scenarios. It is important to mention that each energy source has been sized to satisfy the demand only. Moreover, for the sake of the analysis, it is assumed that the platforms are not decommissioned; so the demand is constant until 2050.

In Figure 19 the results from the analysis based on Tyra and Halfdan are presented. At first, it should be mentioned that floating wind has the same cost assumptions on both platforms thus the LCOE is equal. From an overall perspective, one can notice a defined trend in the cost development of each energy source. This is due to the underlying cost assumptions. Specifically, the gas turbine's LCOE is highly depended on the $CO_2$ tax as it can be seen comparing the trend with the $CO_2$ tax development in Figure 11. Moreover, there is a large difference in the range of the gas turbine LCOEs between the two clusters. The reason behind this lies in the efficiency assumed for the gas turbines on the platforms. Tyra, which has just been renovated, has installed a gas turbine twice as efficient as the gas turbine on Halfdan. This results in a lower fuel consumption which was found to be the largest share of the costs in Figure 14. The LCOE of the delivered electricity involves the capital cost of the transmission lines and the costs of producing electricity, which is the largest share so it influences the cost development overtime.

Comparing the energy sources between each other, a diverging trend between floating wind and the Gas Turbine can be noticed. The Gas Turbine is the most expensive option in all years, except for in 2025 on Tyra. Delivering electricity is the best solution for both clusters until 2045, when floating wind costs become competitive. Moreover, in both cases, floating wind could be installed at the earliest in 2030 when its costs are reduced by 25%. This analysis helps to explain the trends seen in the results in terms of the LCOEs of the competing options.

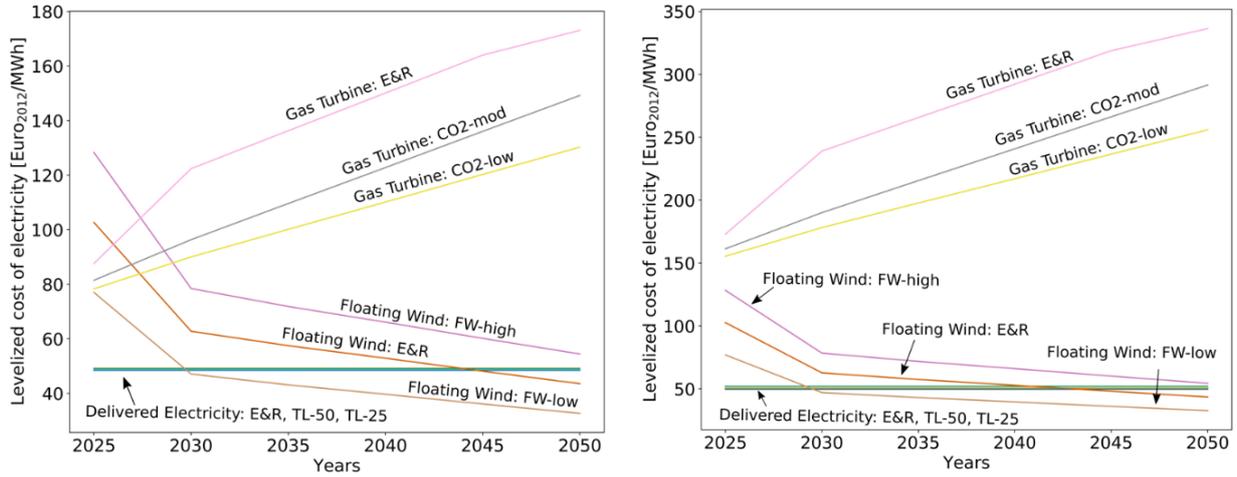

*Figure 19 Levelized Cost of Electricity analysis for Tyra (left) and Halfdan (right)*

## 4.3.4.2. Allocation of hydrogen production for transport across Europe

In Balmorel a demand for the decarbonisation of the transport sector as discussed in section 4.1 was included. The model can allocate the demand in any of the country included in the model up a total capacity. Figure 20 shows a comparison between the allocation of demand among the countries in *BAU* versus *E&R* through the period.

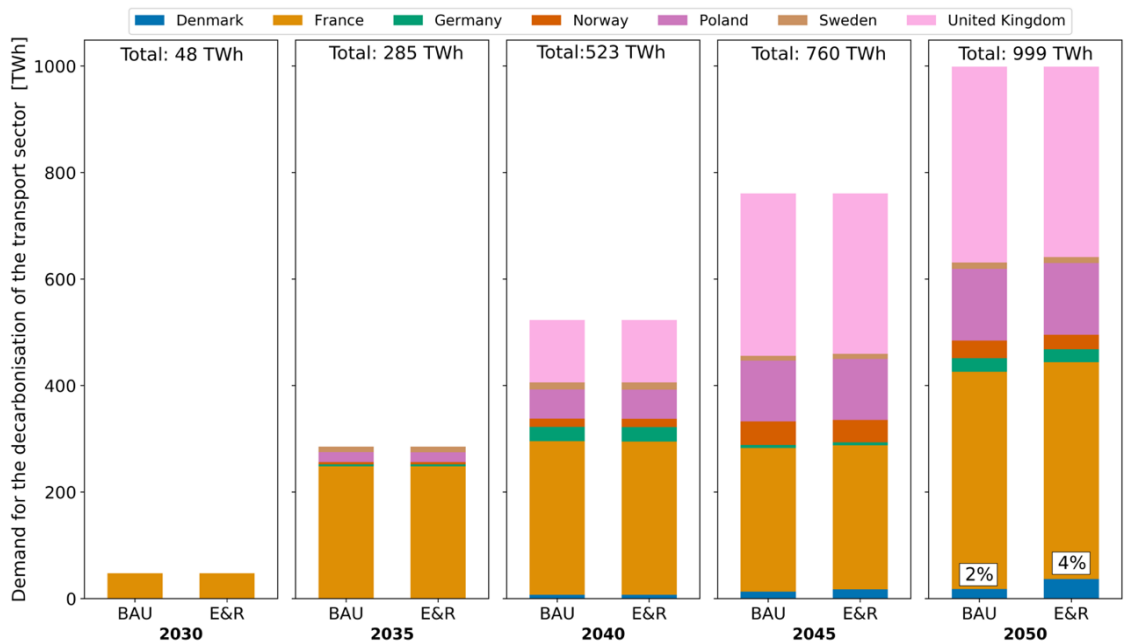

*Figure 20 Allocation of the demand for the decarbonisation of the transport sector among the countries considered in the model in BAU vs E&R*

The total demand for each year is equal in both scenarios since it is an exogenous variable. The demand increases by more than 20 times from 48 TWh in 2025 to about 1000 TWh in 2050. From the figure, one can notice that France is the first country to satisfy the demand and its share is the largest across all years. Furthermore, the demand allocation among the countries is quite similar between the two scenarios except in 2050 where Denmark's share doubles from 2% (18 TWh) to 4% (37 TWh). This is due to the large investments in floating wind which provides electricity at a competitive price for hydrogen production through electrolysis.

### 4.3.4.3. Investments in floating wind and hydrogen plant

A sensitivity analysis was performed to evaluate the drivers for the investment in floating wind and hydrogen plant. The results are presented in Figure 21 and Figure 22. In the figures, for each modeling year (e.g. 2025, 2030…), the cumulative installed capacity and the share of capacity on each cluster is provided for the core scenarios plus the *E&R* scenario as a reference.

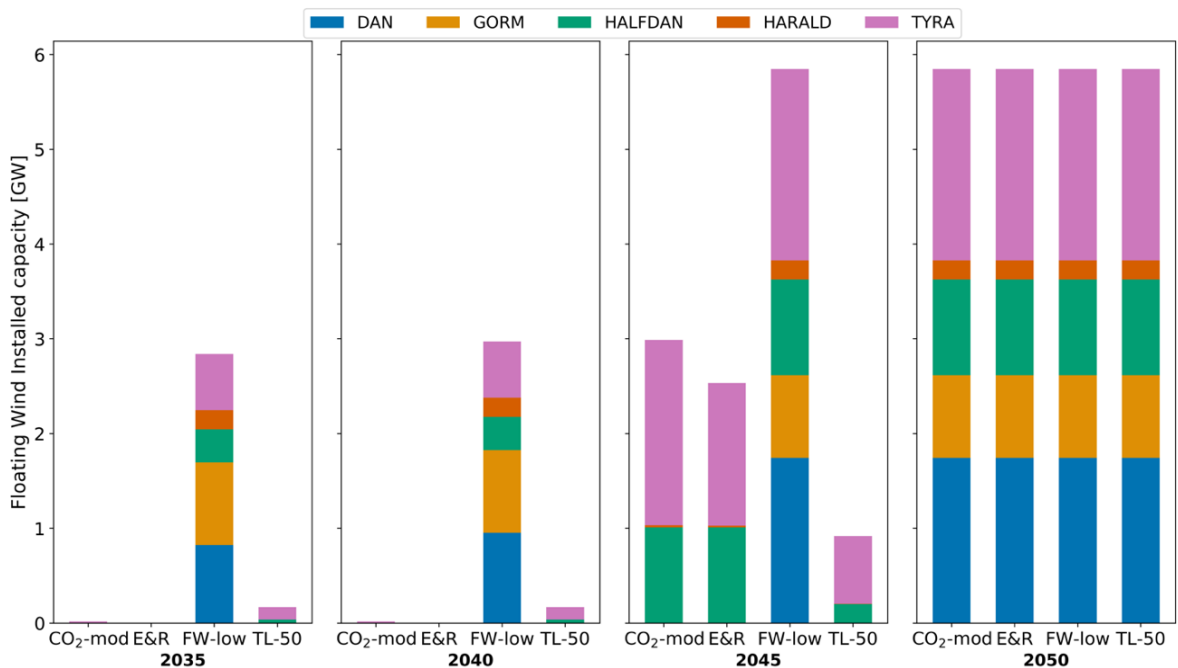

*Figure 21 Floating wind capacity for each modeling year and cluster for the core scenarios*

Comparing the two figures, one can notice a correlation between the installed capacity of the two technologies. Floating wind is a driver for the production of hydrogen offshore. In the *FW-low* scenario, where floating wind costs are reduced by 25%, hydrogen is already produced in 2035. Moreover, the distribution of the technologies' capacity among the clusters highlights the correlation between the two technologies. In 2045, in all scenarios but *FW-low*, Tyra and Halfdan clusters have the largest share of capacity of both technologies.

Focusing on floating wind, it is interesting to notice that in 2050 the cumulative capacity and its distribution among the cluster is the same for all scenarios. Therefore, it can be concluded that floating wind is a very competitive energy source in 2050 while it less competitive earlier in time, except if its costs are decreased as in FW-Low.

On the other hand, the cumulative capacity of the hydrogen plant in 2050 varies between the scenarios. On average about 3 GW are installed across all scenarios.

However, the share of capacity is highly dependent on the scenario. Dan and Tyra account for the largest shares of capacity in 2050, but Dan has the lowest share in 2045 among three scenarios. Moreover, in the *FW-low* scenario, in 2050, Gorm does not produce any hydrogen. From these results, it can be concluded that the production of hydrogen from electricity produced on the platform is not strongly competitive. Therefore, the variations performed on the variables of the sensitivity scenarios have a high influence on the allocation of the electricity for the hydrogen production.

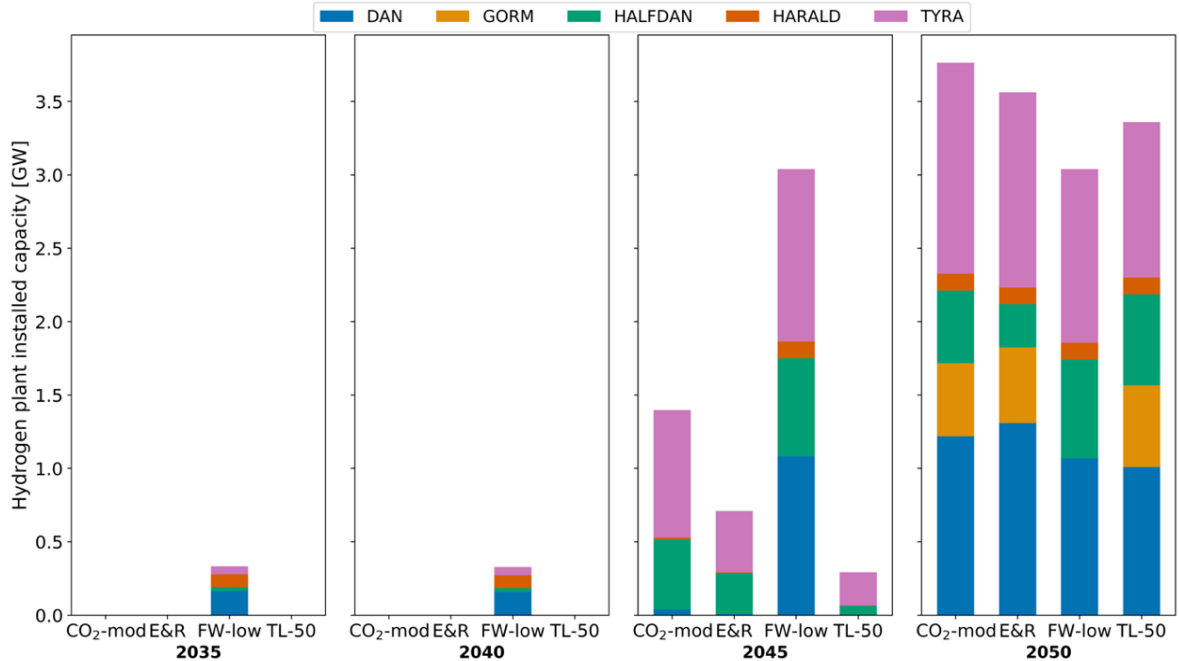

Figure 22 Electrolyser capacity for each modeling year and cluster in the core scenarios.

### 4.3.4.4. LCOEs of hydrogen in 2050 for each platform

The repurposing of the platform for alternative uses involves large investments which can be offset by the sale of the energy produced on the platform. The LCOE shows the price of energy to break-even with the investments. Table 13 shows the hydrogen LCOE in €$_{2012}$/kg H$_2$ that is required to offset the yearly expenses of each cluster in 2050 – minus the income from the sale of electricity. The analysis was performed on all scenarios and for each cluster. The *E&R* scenario represents the reference while the other scenarios were used to identify the range of variations of the LCOE which is provided in the last two columns of Table 13.

Table 13 H$_2$ LCOEs in 2050 for each platform in the E&R scenario. The last two columns show the minimum and maximum LCOE observed in the sensitivity scenarios.

| LCOE [€$_{2012}$ / kg H$_2$] | | | |
|---|---|---|---|
| **Cluster** | **E&R** | **min** | **max** |
| Dan | 3.55 | 3.55 | 4.13 |
| Gorm | 3.95 | 3.17 | 4.09 |
| Halfdan | 6.75 | 3.78 | 59.57 |
| Harald | 4.48 | 3.97 | 4.68 |
| Tyra | 3.94 | 3.35 | 4.50 |

The analysis shows that the LCOE on average is 4.53 €$_{2012}$/kg H$_2$ (4.89 €$_{2020}$/kg H$_2$), which is significantly higher than the expected price of renewable hydrogen of 0.7-1.5 €/kg H$_2$ already in 2030 (Hydrogen Europe 2020). The LCOEs range is relatively small among all clusters except for Halfdan. On this cluster, there is a low income from the electricity sale and a low hydrogen production. The combination of these two conditions increases the LCOE highly.

### 4.3.4.5. Sensitivity analysis of the hydrogen LCOE

A sensitivity analysis on the LCOE of the hydrogen was performed. It is considered that the hydrogen plant is installed in the *E&R* scenario on Tyra in 2050. In the analysis, each cost component of the LCOE was changed by +/- 20% at a time and the variation of the LCOE was recorded. Figure 23 provides the list of cost components and the corresponding positive and negative percentage variation of the LCOE compared to the reference case, 3.94 €$_{2012}$/kg H$_2$.

The sensitivity analysis demonstrated that the electrolyser efficiency has the highest influence on the LCOE with a correlation close to 1. The LCOE reaches 4.71 and 3.16 €$_{2012}$/kg H$_2$ when the negative and positive variation of the electrolyser efficiency, respectively.

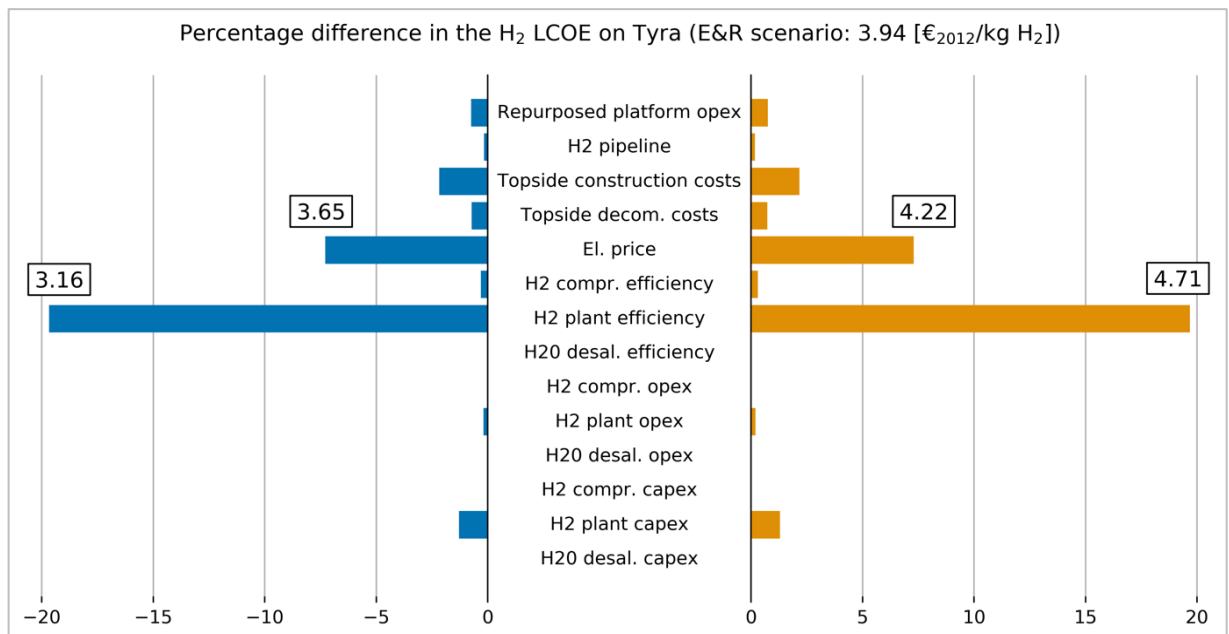

*Figure 23 Sensitivity analysis of the H$_2$ LCOE on Tyra in 2050*

Between the other costs' components, the electricity price has the second highest influence followed by the electrolyser Capex. This is in line with the findings of Hydrogen Europe which estimate the electricity price to account for 60-80% of the hydrogen cost (Brunner et al. 2015; Götz et al. 2016). All the other costs' components results in a variation of the LCOE lower than 2.5%.

It is important to mention that the costs of repurposing and keeping the platform operational are independent from the costs of the hydrogen plant. Therefore, the possible influence of these costs on the hydrogen LCOE cannot be evaluated fully. Furthermore, performing the sensitivity analysis on the Tyra cluster provides useful insights but cannot represent the situation on all platforms. However, we can see that the LCOE is influenced the most strongly by the electrolyser's efficiency which is an exogenous variable equal for all platforms. Therefore, any difference in the sensitivity analysis would results from differences in the electricity prices among the platforms, which are very similar.

## 4.4. Discussion

This section critically discusses the employed results (section 4.4.1) and the method (4.4.2), in this order, before giving some recommendations (4.4.3).

### 4.4.1. Results

The electrification of all O&G platforms in the first years of analysis is an interesting result, especially as the platforms are connected to the shore and energy islands rather than being electrified with floating offshore wind (as in several case studies, cf. section 2.4.1). This electrification results in significant cost and emissions reduction for energy supply to the platforms. Even with substantially higher transmission line costs (i.e. TL-25 and TL-50, see Figure 19), this remains the cheaper option for electrification compared to offshore wind, especially once the offshore energy island with up to 10 GW of connected offshore wind is established by 2030. Electrifying the platforms provides a large degree of flexibility, as this electricity can be used for multiple applications, including motive power, light as well as space and process heating. The connection of these platforms to the Danish and North Sea electricity transmission grid could also prepare them for future alternative uses and provides many opportunities in this regard. Some of these are explored in this study, but others might involve O&M, logistics and/or shipping hubs.

The second major noteworthy result relates to the repurposing of the O&G platforms as an alternative to decommissioning. In all cases, these platforms are connected to up 5.8 GW of floating offshore wind and repurposed with new topsides and up to 3.6 GW of electrolysers to serve as hydrogen generators. Seen in the context of expected total electrolyser capacity of 40 GW in Europe by 2030, which requires the realization of up to 80 GW of additional renewable electricity production, these results seem reasonable (Ad van Wijk & Jorgo Chatzimarkakis, 2020). This hydrogen is assumed to be transported to the Danish shore through existing and/or new purpose-built pipelines and there stored/transported for use in the transport sector. In the context of the system-level analysis carried out in this study, with all North Sea neighbouring countries and more included in the model, the Danish share of overall hydrogen demand in the *E&R* scenario is 37 TWh compared to just 18 TWh in the *BAU scenario*. In other words, the repurposing of the Danish O&G infrastructure is instrumental in providing an extra 2% of the total future demand for hydrogen in the transport sector by 2050 (cf. Figure 20). This means that it is more economical to exploit these offshore wind resources in the North Sea alongside the repurposing of existing O&G infrastructure, compared to utilizing renewable technologies such as wind and solar PV in other locations onshore.

These results are clearly significant, but they should be understood in the wider context of the method employed and assumptions made in this study. For example, whilst the levelized costs (LCOEs) of hydrogen in this analysis in 2050 at around 4 $€_{2012}$/kg $H_2$ lies significantly above other studies and expectations for the hydrogen price within this timeframe. This calculation does not take into account the support policies that renewable energy technologies benefit from, however. For green hydrogen, such policies would most likely be required and could serve to narrow this gap with other studies of about 2 $€_{2012}$/kg $H_2$, if not making it disappear altogether.

Another related point that should be noted, is the lack of a consideration of markets in this analysis. Both for electricity and hydrogen, current and future markets provide a framework within which any business model needs to operate. For electricity, this currently means operating on one or more of a variety of markets for energy (kWh) and or power (kW) at different temporal and spatial scales. In the context of hydrogen generation through electrolysis, optimizing the operation within these different markets – and especially combining the participation on several markets simultaneously, known as stacking – could provide added economic incentives leading to higher full load hours and reduced levelized costs (DEA 2021b). One salient example involves the use of otherwise curtailed or excess renewable electricity to drive the electrolysis process, which was not assessed here. The inclusion of the market dimension in future analyses could therefore substantially improve the case for generating offshore hydrogen by indicated extra revenue streams and reducing the identified range of break-even costs.

Another important result relates to the costs of decommissioning versus repurposing. Whilst the results indicate that the high costs of decommissioning can in most cases be avoided by repurposing the existing infrastructure, in some cases these repurposing costs are actually higher than the decommissioning costs. The difference is compensated by the additional value that the infrastructure has once repurposed, especially (as analysed here) in terms of serving the future low-carbon energy system, but also (not analysed) possibly as a hub for logistics, O&M and/or shipping activities. On the one hand, there are obvious role(s) of offshore O&M infrastructure to the development of an offshore renewable power system. On the other hand, the uncertainty around the costs and benefits of these roles is very high, especially beyond 2030 which is a crucial timeframe for hydrogen in these results. These uncertainties, as well as some other important aspects of the adopted approach, are discussed in the following section.

### 4.4.2. Method

This section highlights some of the main weaknesses and uncertainties in the employed method, starting by addressing economic and technical aspects in turn, before discussing some general points.

Most if not all of the **economic assumptions** employed in this study are subject to varying degrees of uncertainty. In general, there is high confidence about the economic characterisation of current technologies, as employed in the Balmorel model. These assumptions are based on widely-recognized and authoritative sources such as the Danish Energy Agency's Technology Catalogue (DEA 2021b). Larger uncertainties relate to the economic assumptions for the O&G platforms, for which data was largely based on third-party studies from other contexts outside the DUC. This was inevitable given the fact that few Danish platforms have been decommissioned, and there is arguably a strong similarity between the infrastructure in the UK and Danish sectors, for example. But the fact that platform decommissioning in the North Sea in general is still its infancy, means that any economic assumptions used here for decommissioning and repurposing are uncertain. In addition, the platform operational costs are indirect estimates based on the aggregated operational costs for the whole O&G sector, which are provided by the Danish Energy Agency (DEA 2021a). Finally, the uncertainty relating to all of these economic assumptions clearly increases with time into the future, including the gas and $CO_2$ prices as well as the retirement schedules for O&G platforms. For example, the latter assumptions lead to a cease of operation on Gorm in 2025 but hydrogen production first in 2045, which is probably unrealistic. Estimating economic parameters three decades in advance is obviously a very challenging task and the approach taken here overlooks any potential 'shock' impacts on the energy system, such as the ongoing COVID-19 pandemic. For this reason, it should be emphasized that the absolute results contained in this study are not as important as the relative ones – in other words, the differences between the respective scenarios have more significance than the values contained in them.

From a **technical perspective**, this study also has several weaknesses, especially but not only relating to the O&G infrastructure. As mentioned above, the repurposing of the O&G platforms is largely based on third party data from other sector/countries, so also the technical details are uncertain. For example, the space availability on existing/new topsides, the structural integrity of existing jackets and the energy management aspects of the platform were all at least partly overlooked. The latter could be important for meeting both heat and power demand on the platform, whereby the heat demand in this context was completely neglected. In addition, technical challenges relate to the transportation of hydrogen in pipelines to shore. To account for the problem of hydrogen embrittlement in existing/old oil/gas pipelines, the scenario with the new pipelines alongside the one with the existing ones was considered. But other options for transporting this hydrogen to shore, or elsewhere, were overlooked. The same applies to alternative energy carriers such as ammonia, which could be produced directly on the platform and either transported to shore or directly used as a fuel for ships. Finally, the focus on offshore wind was justified based on the advanced development stage of this technology and the good wind resources in the North Sea, which meant excluding less mature technologies such as salinity gradients, wave energy, ocean thermal energy, geothermal energy etc. All of these aspects should be addressed in future work.

Finally, there are some **general aspects of the methodology** that should be mentioned here. Probably most important is the macroeconomic/socioeconomic perspective adopted, i.e. that of the omnipotent central planner, which does not reflect reality with a mixture of operators and assets. The implication of this is that some of the results in this study may not be economically attractive from the perspective of individual operators – whereby this relates less to the electrification and more to the latter repurposing activities. The issue lies in the apportioning of costs between different fields and platforms, which may be influential for the overall economics. It is therefore recommended to analyse the business case for these measures from an operator's microeconomic perspective in further work. In addition, the spatial resolution employed to include the O&G sector in the Balmorel model is relatively low, such that whole fields with several platforms are aggregated into one cluster. This obviously overlooks any connections between the platforms and related space or energy system constraints. Another general limitation with this method is the simplified way in which hydrogen is modelled, indirectly as electricity demand. Whilst this has advantages in terms of modelling simplicity, it does overlook some important aspects of hydrogen demand and competition (especially markets, see above), which could also be decisive for the business case. Despite there not yet being an established market for hydrogen, this is likely to change within the long timeframes considered here. It is therefore also recommended that an integrated energy system analysis, with hydrogen as a separate energy carrier, is carried out in future work. A more holistic analysis could also include other sectors such as marine transport and thereby also assess synergy effects with alternative uses of repurposed offshore O&G infrastructure as refuelling stations, for example.

### 4.4.3. Recommendations

This section briefly summarizes the recommendations resulting from this feasibility study. Given the explorative nature of this research, these recommendations mainly relate to areas where future work should focus rather than concrete actions for implementation:

- **The business case** for the results reported here should be analysed from the operator's perspective, in order to provide a clear indication of possible value opportunities in repurposing existing assets.
- **Collaborative research with offshore O&G engineers** should clarify the technical constraints on electrification and repurposing, especially but not only relating to the space availability, structural integrity, and energy system integration aspects. This analysis would involve assessing challenges and opportunities for repurposing individual, exemplary platforms.
- **A more holistic energy system analysis** should be performed, which is at a higher spatial resolution and thereby includes details of individual platforms (rather than only clusters), includes hydrogen (and possibly other electro-fuels) as a distinct energy carrier and considers demand and markets for this alongside electricity (which is already included), reflects the energy management system on the platforms and the heat demand, and alternative transportation means for the generated fuels as well as alternative use cases such as refuelling stations.
- **A wider policy, regulation and market analysis** needs to build on the above system analysis in order to assess the required framework conditions for a future integrated North Sea energy system, which maximises social utility by providing adequate incentives for operators and investors. This analysis would provide clear recommendations for national and international policymakers relating to the future development of existing and new North Sea energy infrastructure.

# 5. Summary and conclusions

This study sets out to analyze future synergies between the O&G and renewables sectors and explore how exploiting these synergies could lead to economic and environmental benefits. By firstly reviewing and highlighting relevant technologies and related projects, this report synthesizes the state of the art in offshore energy system integration, by focusing on three key technologies in electrification/floating wind, electrolysis/hydrogen production and Carbon Capture and Storage (CCS). Both existing O&G assets and these renewable technologies are techno-economically analysed with relevant technical and economic criteria. This results in an overview of the interconnected system of O&G assets within the DUC along with operational data and retirement schedules, as well as techno-economic characteristics of floating offshore wind, electrolysis and CCS technologies.

All of these preliminary results serve as input data for a holistic energy system analysis in the Balmorel modelling framework. With a timeframe out to 2050 and model scope including all North Sea neighbouring countries, this analysis explores a total of nine future scenarios for the North Sea energy system. The main results include an immediate electrification of all operational DUC platforms by linking them to the shore and/or a planned Danish energy island. These measures result in cost and $CO_2$ emissions savings compared to a BAU scenario of 72% and 85% respectively. When these platforms cease production, this is followed by the repurposing of the platforms into hydrogen generators with up to 3.6 GW of electrolysers and the development of up to 5.8 GW of floating wind. The generated hydrogen is assumed to power the future transport sector, and is delivered to shore in existing and/or new purpose-built pipelines. The contribution of the O&G sector to this hydrogen production amounts to around 19 TWh, which represents about 2% of total European hydrogen demand for transport in 2050. The levelized costs (LCOE) of producing this hydrogen in 2050 are around 4 €$_{2012}$/kg $H_2$, which is around twice those expected in similar studies. But this does not account for energy policies that may incentivize green hydrogen production in the future, which would serve to reduce this LCOE to a level that is more competitive with other sources.

The very short timeframe and small project team mean that the analysis presented here is relatively high-level, requiring many simplifying assumptions and resulting in some important technical details being overlooked and left for future research. In particular there remain significant uncertainties relating to the technical feasibility and future economic developments of most of the technologies analysed. For this reason, this report should be understood as a starting point for further and more detailed analysis, rather than a definitive roadmap for the sector. Given the explorative nature of this research, recommendations mainly relate to areas where future work should focus:

- The business case for the results reported here should be analysed from the operator's perspective, in order to provide a clear indication of possible value opportunities in repurposing existing assets.
- Collaborative research with offshore O&G engineers should clarify the technical constraints on electrification and repurposing.
- A more holistic energy system analysis should be performed in order to overcome some limitations in spatial resolution and technical approximations.
- A wider policy, regulation and market analysis needs to build on the above system analysis in order to assess the required framework conditions for a future integrated North Sea energy system.

# 7. Appendix

*Table 14 Techno-economic assumptions of the electrolyser plant in 2050 used in the Balmorel*

| | Capacity [kW] | Capex [$€_{2012}$/kW] | Opex fixed [% of Capex] | Opex variable [$€_{2012}$/ $kgH_2$] | Efficiency [kWhel / $kgH_2$] |
|---|---|---|---|---|---|
| Electrolyser | 200 000 | 274.4 | 2 | 1.43 | 43.24 |
| Water Desalinator | 2 | 0.5 | 8 | 0.0016 | 0.048 |
| Hydrogen Compressor | 3000 | 7.0 | 3 | 0.02 | 0.67 |
| | Capacity [GW] | Capex [k€/GW/km] | | | |
| Hydrogen Pipeline 36" (Global CCS institute 2020) | 10 | 179 | | | |
| Reuse of existing gas pipeline | | 17,9 | | | |

*Table 15 Yearly energy related expenses in M€2012 for each platform in the BAU scenario*

| Cluster | Year | $CO_2$ tax | OPEX fixed costs | Fuel costs | OPEX variable costs |
|---|---|---|---|---|---|
| Dan | 2025 | 9.8 | 0.7 | 45.9 | 1.4 |
| | 2030 | 24.8 | 0.7 | 49.9 | 1.4 |
| | 2035 | 29.7 | 0.7 | 52.8 | 1.4 |
| Gorm | 2025 | 5.0 | 0.4 | 23.6 | 0.7 |
| Halfdan | 2025 | 4.6 | 0.3 | 21.7 | 0.7 |
| | 2030 | 11.7 | 0.3 | 23.6 | 0.7 |
| | 2035 | 14.1 | 0.3 | 25.0 | 0.7 |
| | 2040 | 16.4 | 0.3 | 26.4 | 0.7 |
| Harald | 2025 | 0.9 | 0.1 | 4.3 | 0.1 |
| | 2030 | 2.3 | 0.1 | 4.7 | 0.1 |
| Tyra | 2025 | 10.3 | 0.7 | 48.0 | 1.5 |
| | 2030 | 26.0 | 0.7 | 52.3 | 1.5 |
| | 2035 | 31.2 | 0.7 | 55.3 | 1.5 |
| | 2040 | 36.3 | 0.7 | 58.4 | 1.5 |

*Table 16 Cost and $CO_2$ emissions savings in 2025 for each cluster of platforms.*

| Cluster | Costs (BAU) [Mill€$_{2012}$/year] | Costs (E&R) [Mill€$_{2012}$/year] | $CO_2$ emissions (BAU) [ktons] | $CO_2$ emissions (E&R) [ktons] | Share of Denmark's total $CO_2$ emissions (BAU) [%] | Share of Denmark's total $CO_2$ emissions (E&R) [%] |
|---|---|---|---|---|---|---|
| Dan | 59 | 14 | 368 | 39 | 4.42 | 0.5 |
| Gorm | 32 | 7 | 244 | 75 | 2.98 | 0.9 |
| Tyra | 28 | 6 | 169 | 14 | 2.08 | 0.2 |
| Harald | 6 | 1 | 36 | 5 | 0.45 | 0.1 |
| Halfdan | 62 | 28 | 388 | 43 | 4.65 | 0.5 |
| **Total** | **186** | **57** | **1206** | **176** | **14.58** | **2.2** |

*Table 17 Annualized costs in M€$_{2012}$/year for each cluster in 2050*

| Cluster | FW costs | H$_2$ plant Capex | H$_2$ plant Opex | H$_2$ pipeline Costs | Topside decom. costs | Repurposed Platform Opex | Topside construction costs | Electricity Costs | Interconnection Costs | Total |
|---|---|---|---|---|---|---|---|---|---|---|
| Dan | 222 | 503 | 277 | 6 | 24 | 24 | 76 | 49 | 7 | 1189 |
| Gorm | 111 | 198 | 109 | 4 | 19 | 20 | 62 | 14 | 3 | 540 |
| Halfdan | 140 | 113 | 61 | 1 | 14 | 15 | 36 | 3 | 4 | 388 |
| Harald | 26 | 44 | 24 | 0 | 5 | 5 | 13 | 5 | 1 | 123 |
| Tyra | 275 | 511 | 278 | 6 | 27 | 28 | 80 | 37 | 9 | 1250 |
| **Total** | **775** | **1369** | **749** | **18** | **89** | **92** | **268** | **107** | **23** | **3490** |

*Table 18 Floating wind farm capacity in GW per cluster, scenario and year*

| Year | Scenario | Dan | Gorm | Halfdan | Harald | Tyra | Total |
|---|---|---|---|---|---|---|---|
| 2035 | CO2-mod | | | | | 0.02 | 0.02 |
| | E&R | | | | | | |
| | FW-low | 0,8 | 0,9 | 0.3 | 0.2 | 0.6 | 2.8 |
| | TL-50 | | | 0.0 | | 0.1 | 0.2 |
| 2040 | CO2-mod | | | | | 0.02 | 0.02 |
| | E&R | | | | | | |
| | FW-low | 1.0 | 0.9 | 0.3 | 0.2 | 0.6 | 3.0 |
| | TL-50 | | | 0.04 | | 0.1 | 0.2 |
| 2045 | CO2-mod | | | 1.0 | 0.02 | 2.0 | 3.0 |
| | E&R | | | 1.0 | 0.02 | 1.5 | 2.5 |
| | FW-low | 1.7 | 0.9 | 1.0 | 0.2 | 2.0 | 5.8 |
| | TL-50 | | | 0.2 | 0.003 | 0.7 | 0.9 |
| 2050 | CO2-mod | 1.7 | 0.9 | 1.0 | 0.2 | 2.0 | 5.8 |
| | E&R | 1.7 | 0.9 | 1.0 | 0.2 | 2.0 | 5.8 |
| | FW-low | 1.7 | 0.9 | 1.0 | 0.2 | 2.0 | 5.8 |
| | TL-50 | 1.7 | 0.9 | 1.0 | 0.2 | 2.0 | 5.8 |

*Table 19 Hydrogen plant capacity in GWel per cluster, scenario and year*

| Year | Scenario | Dan | Gorm | Halfdan | Harald | Tyra | Total |
|------|----------|-----|------|---------|--------|------|-------|
| 2035 | CO2-mod |  |  |  |  |  |  |
|      | E&R |  |  |  |  |  |  |
|      | FW-low | 0.16 |  | 0.03 | 0.09 | 0.06 | 0.33 |
|      | TL-50 |  |  |  |  |  |  |
| 2040 | CO2-mod |  |  |  |  |  |  |
|      | E&R |  |  |  |  |  |  |
|      | FW-low | 0.2 |  | 0.03 | 0.09 | 0.06 | 0.33 |
|      | TL-50 |  |  |  |  |  |  |
| 2045 | CO2-mod | 0.04 |  | 0.48 | 0.01 | 0.87 | 1.40 |
|      | E&R | 0.01 |  | 0.28 | 0.01 | 0.42 | 0.71 |
|      | FW-low | 1.08 |  | 0.67 | 0.11 | 1.17 | 3.04 |
|      | TL-50 |  |  | 0.06 | 0.001 | 0.23 | 0.29 |
| 2050 | CO2-mod | 1.22 | 0.50 | 0.49 | 0.12 | 1.44 | 3.76 |
|      | E&R | 1.31 | 0.52 | 0.29 | 0.11 | 1.33 | 3.56 |
|      | FW-low | 1.07 |  | 0.67 | 0.12 | 1.18 | 3.04 |
|      | TL-50 | 1.01 | 0.56 | 0.62 | 0.11 | 1.06 | 3.36 |

*Table 20 Hydrogen LCOE €$_{2012}$ / kg H$_2$ in 2050 per cluster and sensitivity scenario*

| Cluster | E&R | CO2-low | CO2-mod | FW-low | H$_2$-low | TL-25 | TL-50 |
|---------|-----|---------|---------|--------|-----------|-------|-------|
| Dan | 3.55 | 3.63 | 3.69 | 3.87 | 4.01 | 3.61 | 4.13 |
| Gorm | 3.95 | 3.88 | 4.09 | - | 3.17 | 3.93 | 4.01 |
| Halfdan | 6.75 | 5.78 | 4.73 | 3.78 | 59.57 | 6.50 | 4.10 |
| Harald | 4.48 | 4.50 | 4.47 | 4.68 | 3.97 | 4.45 | 4.55 |
| Tyra | 3.94 | 3.97 | 3.81 | 4.08 | 3.35 | 4.36 | 4.50 |